\documentclass[aps,prb,twocolumn, superscriptaddress]
{revtex4-1}

\usepackage{amsmath} 
\usepackage{amsthm} 
\usepackage{graphicx} 
\usepackage{float}
\usepackage{amssymb}

\begin{document}

\title{Nonlinear Meissner effect in Nb$_3$Sn coplanar resonators.}

\author{J. Makita$^{1}$, C. Sundahl$^{2}$, G. Ciovati$^{1, 3}$, C.B. Eom$^{2}$, and A. Gurevich}

\affiliation{Department of Physics and Center for Accelerator Science, Old Dominion University, Norfolk, Virginia 23528, USA \\
$^{2}$Department of Materials Science and Engineering, University of Wisconsin-Madison, Madison, Wisconsin 53706, USA \\
$^{3}$ Thomas Jefferson National Accelerator Facility, Newport News, Virginia 23606, USA}

\date{\today}%

\begin{abstract}

We investigated the nonlinear Meissner effect (NLME) in Nb$_3$Sn thin film coplanar resonators by measuring the resonance frequency as a function of a parallel magnetic field at different temperatures. We used low rf power probing in films thinner than the London penetration depth $\lambda(B)$ to  significantly increase the field onset of vortex penetration and measure the NLME under equilibrium conditions. Contrary to the conventional quadratic increase of $\lambda(B)$ with $B$ expected in s-wave superconductors, we observed a nearly linear increase of the penetration depth with $B$.  We concluded that this behavior of $\lambda(B)$ is due to weak linked grain boundaries in our polycrystalline Nb$_3$Sn films, which can mimic the NLME expected in a clean d-wave superconductor.
\end{abstract}

\pacs{74.25.-q, 74.25.Ha, 74.25.Op, 74.78.Na}

\maketitle

The Meissner effect is one of the fundamental manifestations of the macroscopic phase coherence of a superconducting state. Meissner screening current density ${\bf J}=-en_s{\bf v}_s$ induced by a weak magnetic field is proportional to the velocity $\bf{v}_s$ of the condensate. At higher fields, the superfluid density $n_s$ becomes dependent on ${\bf v}_s$ due to pairbreaking effects, resulting in the nonlinear Meissner effect (NLME) \cite{GL,YS1,YS2, Dahm1, Dahm2,H1,H2,Prozorov}. For a single band isotropic s-wave superconductor, the NLME is described by:
\begin{equation}
    {\bf J}=-\frac{\phi_0{\bf Q}}{2\pi\mu_0\lambda^2}(1-\Upsilon\xi^2 Q^2),
    \label{JQ} \\
\end{equation}
where $\lambda$ is the London penetration depth, $\xi$ is the coherence length, ${\bf Q}= m{\bf v}_s/\hbar=\nabla\chi+2\pi{\bf A}/\phi_0$, $m$ is the quasiparticle mass, $\chi$ is the phase of the order parameter $\Psi=\Delta e^{i\chi}$, ${\bf A}$ is the vector potential, $\phi_0$ is the magnetic flux quantum, and the factor $\Upsilon(T,l_i)$ depends on the temperature $T$, the mean free path $l_i$ and details of pairing mechanisms.  Ginzburg and Landau (GL) were the first who obtained the field dependent-correction to the penetration depth $\lambda(B)$ of the magnetic field ${\bf B}$ parallel to a semi-infinite superconductor \cite{GL}:  
\begin{equation}
\lambda(B)=\left[1+\frac{\kappa(\kappa+2^{3/2})B^2}{8(\kappa+2^{1/2})^2B_c^2}\right]\lambda,  
\label{gl}
\end{equation}
where $B_c=\phi_0/2^{3/2}\pi\lambda\xi$ is the thermodynamic critical field and $\kappa=\lambda/\xi$ is the  GL parameter.

In recent years, the NLME has attracted much attention as a probe of unconventional pairing symmetries of moving condensates. Particularly, Yip and Sauls ~\cite{YS1,YS2} showed that in a clean  d-wave superconductor at $k_BT<p_Fv_s$ the supercurrent acquires a nonlinear singular term $\propto |{\bf Q}|{\bf Q}$ strikingly different from that in Eq. (\ref{JQ}), where $p_F$ is the Fermi momentum. Yet Eq.~(\ref{JQ}) can describe a variety of nonlinear electromagnetic responses, both in conventional and unconventional superconductors. For instance Eq. (\ref{JQ}) describes a clean d-wave superconductor at high temperatures $T>p_Fv_s/k_B$ or a d-wave superconductor with impurities ~\cite{YS1,YS2,Dahm1,Dahm2,H1,H2}.   In a clean s-wave superconductor, the NLME is absent at $T\ll T_c$ as $\Upsilon\propto\exp(-\Delta/k_BT)$ ~\cite{Bardeen}, but occurs in the dirty limit $(\xi\lesssim l)$ in which $\Upsilon\sim 1$ even at $T\to 0$ ~\cite{Maki}. In multiband superconductors the NLME could probe the proliferation of interband phase textures ~\cite{GV} or the line nodes and interband sign change in the order parameter or mixed $s-d$ pairing symmetries in iron pnictides ~\cite{FBS}.

So far the observations of the NLME in high-$T_c$ cuprates have been inconclusive ~\cite{nme1,nme2,nme3,nme4,nme5,oates,carr}, mostly because of a small field region of the Meissner state in high-$\kappa$ type-II superconductors and contributions of extrinsic materials factors, such as grain boundaries or local nonstoichiometry. Since the NLME becomes essential in fields $B$ of the order of $B_c$, penetration of vortices above the lower critical field $B_{c1}=(\phi_0/4\pi\lambda^2)(\ln\kappa+0.5)\ll B_c$ ~\cite{ehb} limits the nonlinear correction in Eq.~(\ref{gl}) to $(B_{c1}/B_c)^2/8\simeq \ln\kappa/16\kappa^2\ll 1$. Yet even small NLME terms in Eq.~(\ref{JQ}) causes  intermodulation effects \cite{Dahm1,Dahm2} under strong ac fields, as it was observed in YBa$_2$Cu$_3$O$_{7-x}$ ~\cite{nme5,oates}. 

In this paper we investigate the NLME in a thin film Nb$_3$Sn coplanar resonator in a parallel dc magnetic field, which mitigates the problems of vortex penetration and nonequilibrium effects. We used the method of Ref. \onlinecite{ggc} in which the resonant frequency $\omega=(LC)^{-1/2}$ of a coplanar resonator is measured as a function of a parallel dc field $B$. Here $C$ is the strip-to-ground capacitance and $L=L_{g}+L_{k}$ is the total inductance containing both the geometrical inductance $L_g$, and the field-dependent kinetic inductance of the superconducting condensate, $L_{k}(B)\propto \lambda^2(B)$ ~\cite{OD}. To extend the field region of the NLME, we performed our measurements on thin films of thickness $d<\lambda $ for which $B_{c1}=(2\phi_{0}/\pi d^{2})\ln (d/\xi)$ can be much higher than the bulk $B_{c1}$ ~\cite{Abrikos}. Rotating the field in the plane of the film gives rise to an orientational dependence of $\lambda(B)$ ~\cite{ggc}. Unlike the standard quadratic field correction to $\lambda(B)$ observed on Nb ~\cite{ggc}  and Al ~\cite{Al} thin film resonators, we observed a nearly linear field dependence of $\lambda(B)$ in polycrystalline Nb$_3$Sn films. The fact that the NLME in the s-wave superconductor Nb$_3$Sn exhibits the behavior expected from a clean d-wave superconductor \cite{YS1,YS2} shows the importance of materials factors, particularly local nontoichiometry and weakly coupled grain boundaries characteristic of Nb$_3$Sn, cuprates and pnictides ~\cite{gb,grb}. 

The paper is organized as follows. In Sec. II we describe the experimental setup and Nb$_3$Sn coplanar resonator used in the measurements of NLME. Sec. III contains the main experimental results. Sec. IV summarizes the essential mechanisms of NLME and theoretical results necessary for the comparison of theory with experiment.  Sec. V contains discussion of our results.

\section{Experimental}
\subsection{Film Deposition and Patterning}
\begin{figure}[ht]
	\includegraphics[width=1.0\linewidth]{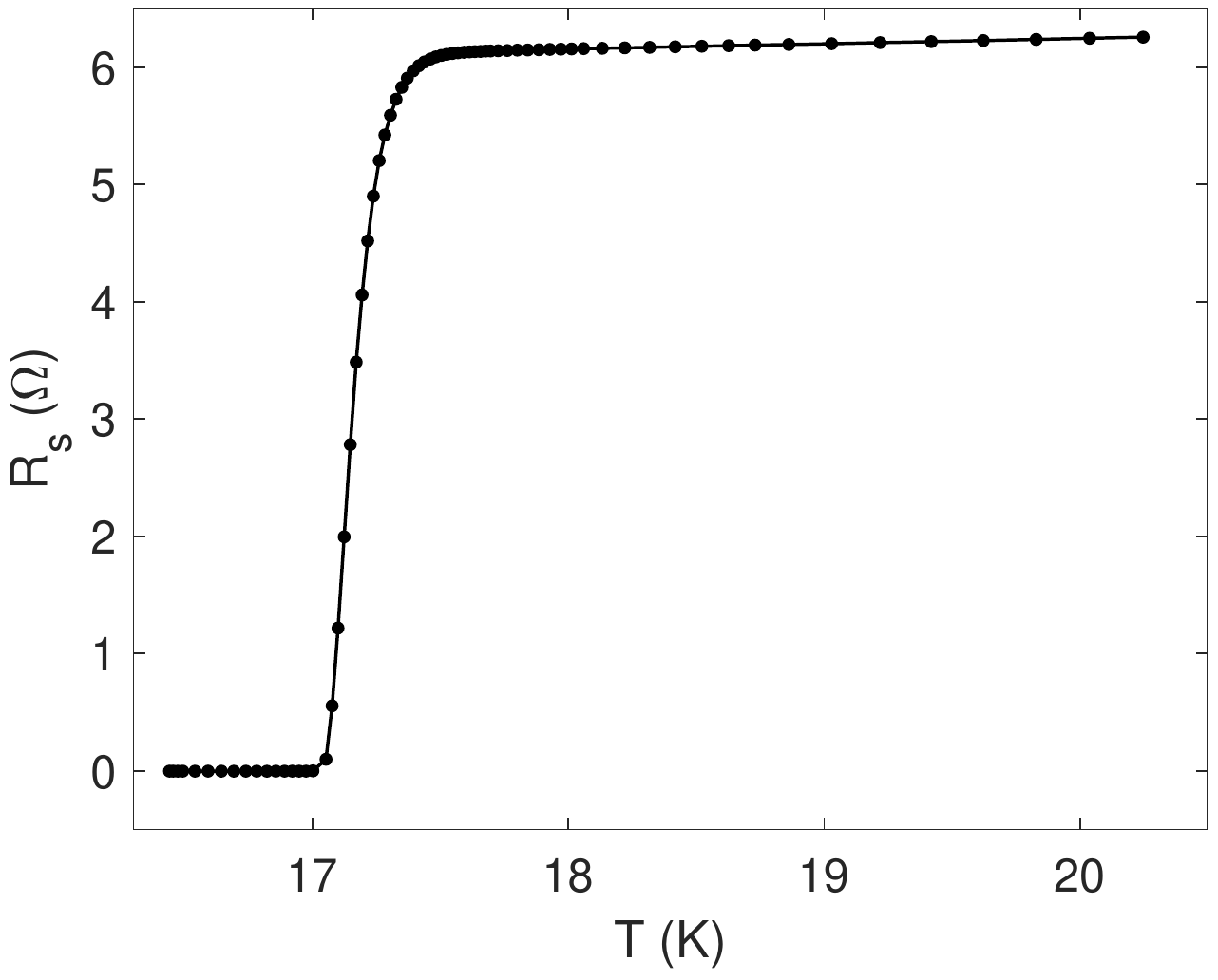}
	\caption{Temperature dependence of the resistance of the film $R_s(T)$ with a midpoint critical temperature $T_c=17.2$ K.}
	\label{fig:rrr}
\end{figure}

The coplanar resonator was fabricated from a 50 nm thick Nb$_3$Sn film on a 10 mm $\times$ 10 mm $\times$ 1 mm Al$_2$O$_3$ substrate. The film was prepared with magnetron co-sputtering using both Nb and Sn targets in a growth chamber at University of Wisconsin-Madision, as described in Refs. \onlinecite{chris,ml}. Figure \ref{fig:rrr} shows the resistive transition in a film grown under a similar condition. The film had a midpoint $T_c\approx 17.2$ K, normal state sheet resistance of $5.1\,\Omega$, and a residual resistance ratio (RRR), $R_s(300\,K)/R_s(18\,K)\approx 3.2$. 

\begin{figure}[ht]
	\includegraphics[width=1.0\linewidth]{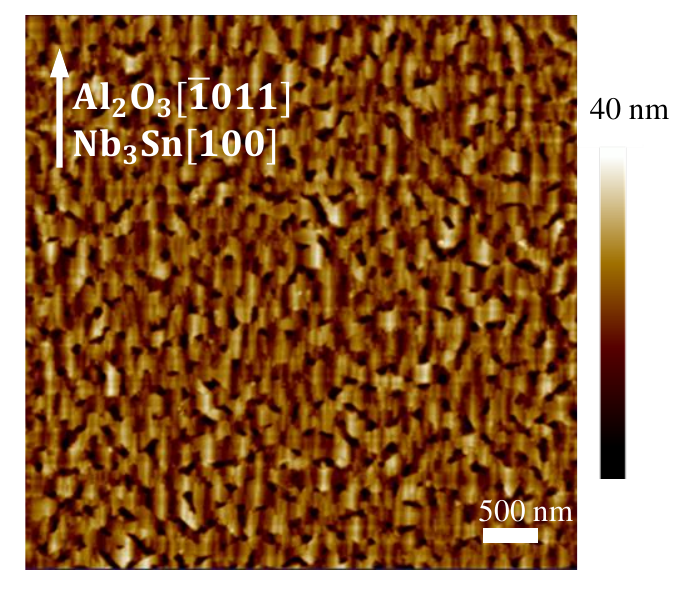}
	\caption{AFM image showing a polycrystalline structure of our films.}
	\label{fig:afm}
\end{figure}

The film has a polycrystalline structure with rigid grains along the [-1011] direction of the Al$_2$O$_3$ substrate as revealed by the atomic force microscopy shown in Fig. \ref{fig:afm}. Those grains contributed to an RMS roughness of approximately 10 nm~\cite{chris,ml}. 

\begin{figure}[ht]
	\includegraphics[width = 1.0\linewidth]{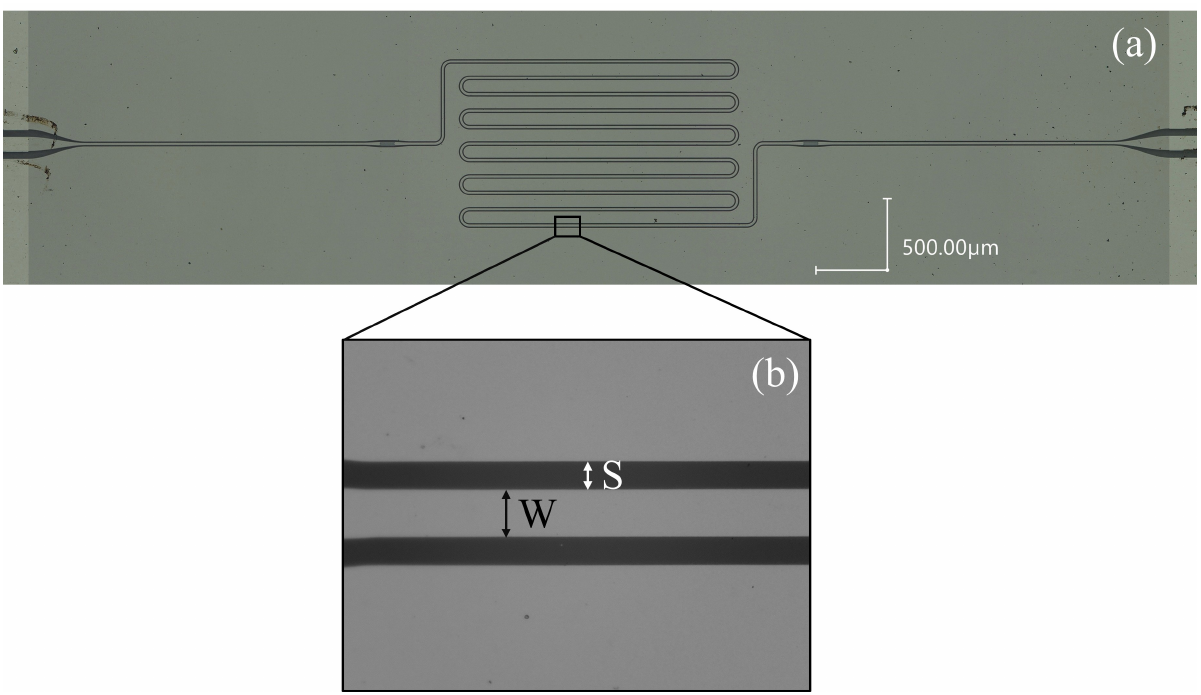}
	\caption{(a) The image of the Nb$_3$Sn coplanar half-wave resonator with $f_0 = 2.236$ GHz. The meandered resonator in the center is terminated capacitively on both ends which tapers out to the input and output landing pads. (b) Zoomed in section of the coplanar resonator where the width of the strip is $w = 15\,\mu\mathrm{m}$ and the gap between the signal strip and the ground is $s = 8.8\,\mu\mathrm{m}$.}
	\label{fig:resonator}
\end{figure}

The sample was patterned into a half-wave coplanar waveguide resonator using contact lithography followed by Ar ion milling. The optical image of the resonator is shown in Fig. \ref{fig:resonator}. The meandered resonator has a total length $l\approx 24.6$ mm corresponding to the fundamental resonant frequency $f_0=2.236$ GHz. The center conductor has a width $w =15\,\mu$m, and a gap width $s=8.8\,\mu$m between the center strip and the ground plane. The $s/w$ ratio was set to achieve a characteristic line impedance $50\,\Omega$. The resonator is coupled to input and output RF probes by interdigital capacitors patterned on the strip.  At the ends of the transmission line, landings pads for ground-signal-ground (GSG) probes were fabricated, shown as the lightly shaded region in Fig. \ref{fig:resonator}(a).  These were made by first removing few nanometers of oxide layers on the surface of Nb$_3$Sn using Ar ion milling and then depositing a 20 nm thick layer of Pd in-situ using a lift-off technique. The landing pads made of Pd serve to prevent oxidation and damage of the film from repeated touchdown of the probes and ensure ohmic contact between the probe and the sample.

\subsection{Measurement Setup} 
\begin{figure}[ht]
	\includegraphics[width=1.0\linewidth]{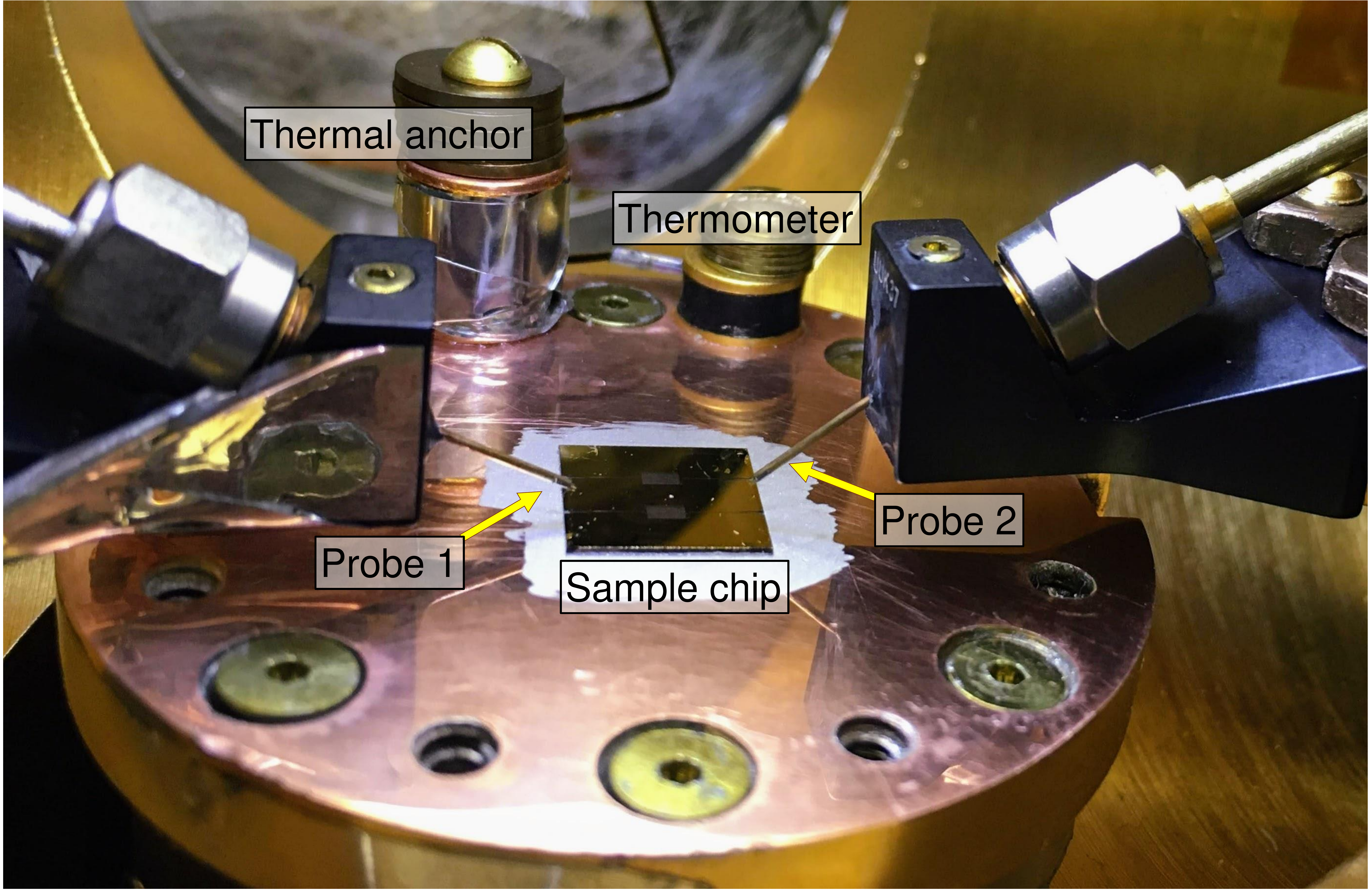}
		\caption{A setup of the sample stage with two GSG probes. The sample is mounted using silver paint, and the thermometer and the OFHC 		bobbin for thermal anchoring of the lead wires are screwed onto the sample stage. }
		\label{fig:samplestage}
\end{figure}

\begin{figure}[ht]
	\includegraphics[width=1.0\linewidth]{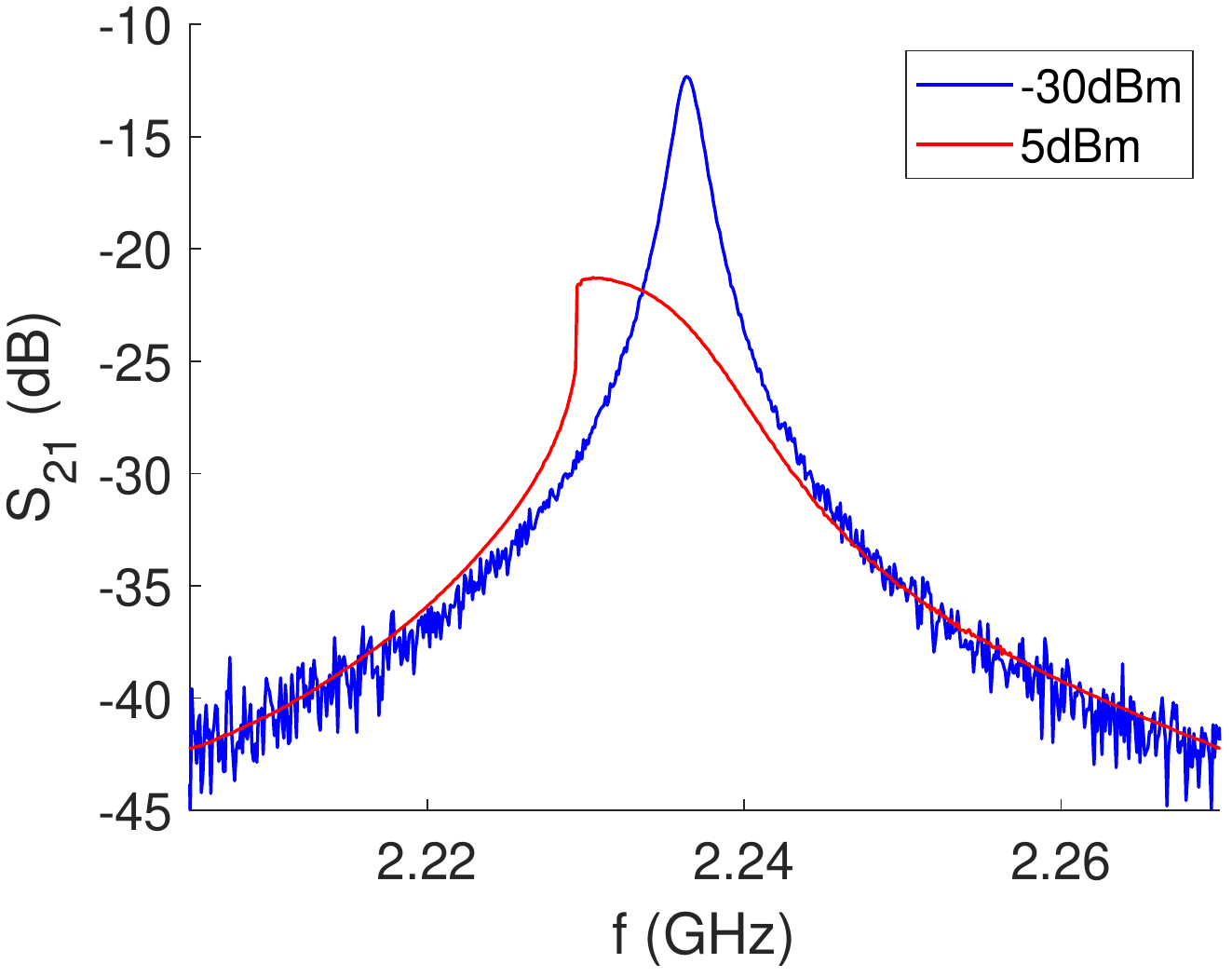}
	\caption{A typical transmission spectrum near resonance displays a symmetric Lorentzian line shape with VNA power output of -30 dBm. Higher input power distorts the Lorentzian shape at -5 dBm.}
	\label{fig:pdependence}
\end{figure}

The patterned sample was mounted inside a cryogenic probe station equipped with a closed cycle cryocooler ~\cite{probe}, as shown in Fig. \ref{fig:samplestage}.  The complex transmission coefficient  $S_{21,12}$ was measured as functions of temperature and the external in-plane magnetic field $B$. The resonant frequencies and the London penetration depth were extracted by fitting the dependence of the phase on frequency of the transmission spectra~\cite{pp,anlage,porch,SL,collin,hein,mohebbi}.  The temperature of the sample was varied using a resistive heater underneath the sample stage, and a parallel dc field up to 200 mT was produced with a NbTi superconducting magnet. This magnet was mounted on a six-motor hexapod system that allowed for fine tuning of magnet orientation by $\pm 7^\circ$ in three axes while taking the sample measurements. The temperature of the sample was measured using a calibrated Cernox (CX-1050-CU-HT, Lakeshore Cryotronics) resistance-temperature device fixed on the stage next to the sample. The output port of a Vector Network Analyzer (VNA) provided rf power that was delivered to the resonator by landing two GSG probes to the contacts.  These probes and the cables connecting to the network analyzer were calibrated using a Short-Open-Load-Through calibration substrate mounted on the sample stage at 7 K. The drive power of VNA was selected to be -30 dBm to maximize the signal-to-noise ratio while avoiding distortion of the Lorentzian shape in transmission signal observed at higher power due to nonlinear heating effects ~\cite{SL,abdo} as shown in Fig. \ref{fig:pdependence}.  To minimize the number of vortices trapped in the sample during its cooldown through $T_c$, we used three pairs of Helmholtz coils to reduce the ambient field $B_a$. The magnitude of $B_a$ was measured by a magnetometer while adjusting the coil currents to achieve $B_a<$4 mG in an optimum configuration. 

The resonance frequency $f_0=(CL)^{-1/2}/2\pi$ is determined by the ground capacitance $C$ and the resonator inductance $L$. Here $L=L_g+L_k$ contains a geometrical inductance $L_g$ and a kinetic inductance $L_k(T,B)$ associated with the inertia of supercurrents.  The geometric inductance for the parameters of our sample $L_g =420.5$ nH/m was calculated in Appendix \ref{A} following Ref. \onlinecite{collin}. The kinetic inductance for a thin film strip of thickness $d < \lambda(T)$ and width $w$ is given by ~\cite{OD,porch,anlage}
\begin{equation} 
L_k\approx \frac{\mu_0\lambda(T)}{w}\coth\left[\frac{d}{\lambda(T)}\right]\approx\frac{\mu_0\lambda^2(T)}{wd}.
\label{Lk}
\end{equation}
Temperature dependencies of $L_k(T)$ and $\lambda(T)$ were inferred from the measured frequency shift $\delta f/f_0=[f_0(T)-f_0(7K)]/f_0(7K)$:  
\begin{equation}
		\frac{\delta f(T)}{f_0(7K)}= \frac{\sqrt{L_g + L_k(7K)}}{\sqrt{L_g+L_k(T)}}-1, 
		\label{dffT}
\end{equation} 
where $C$ and $L_g$ are assumed independent of $T$. By fitting the observed $\delta f(T)/f_0(7K)$ to Eqs. (\ref{Lk}) and (\ref{dffT}), we obtained $\lambda(T)$ as described in the next subsection. 

\begin{figure}[!htb]
		\includegraphics[width=1.0\linewidth]{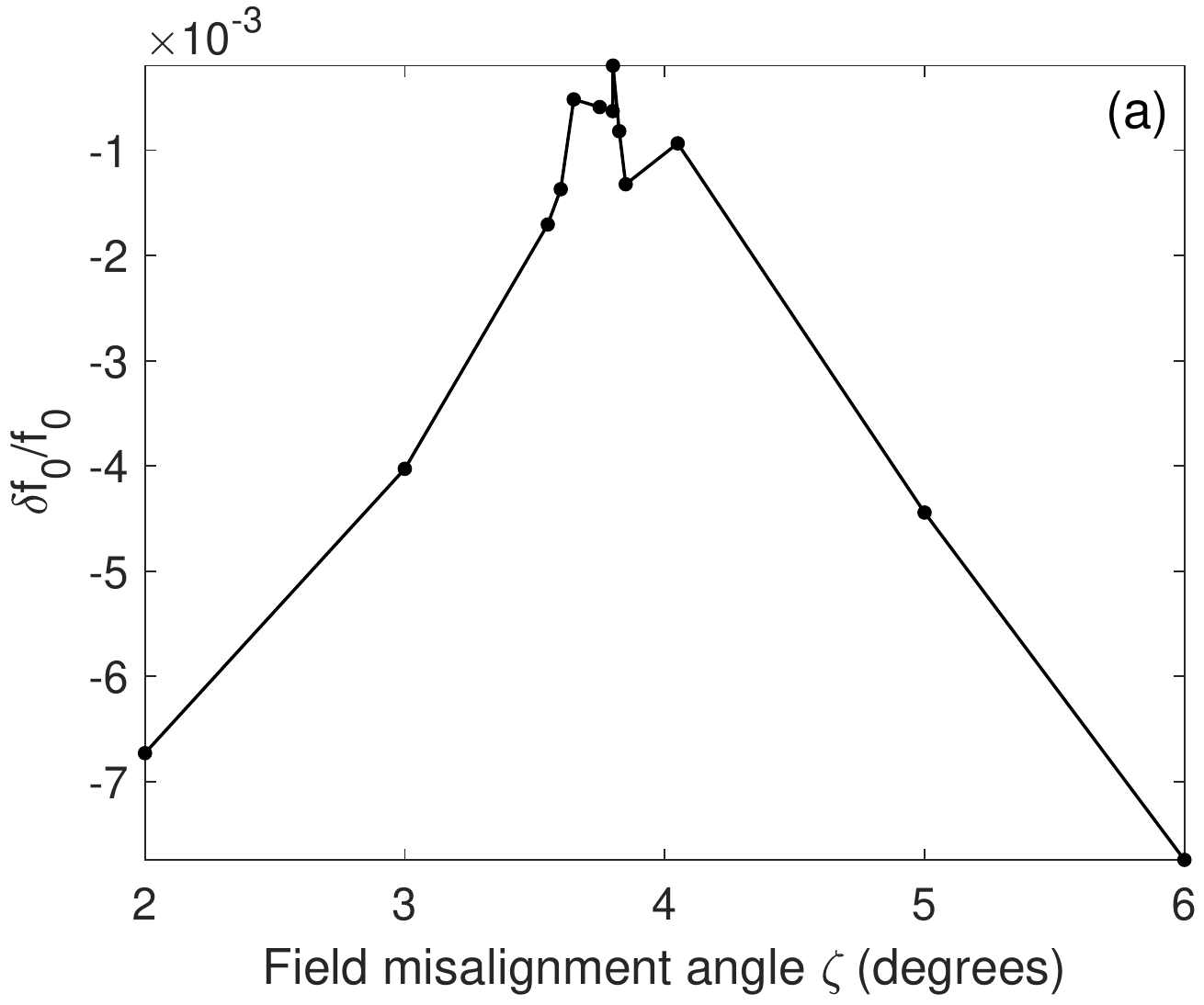}\\
		\includegraphics[width=1.0\linewidth]{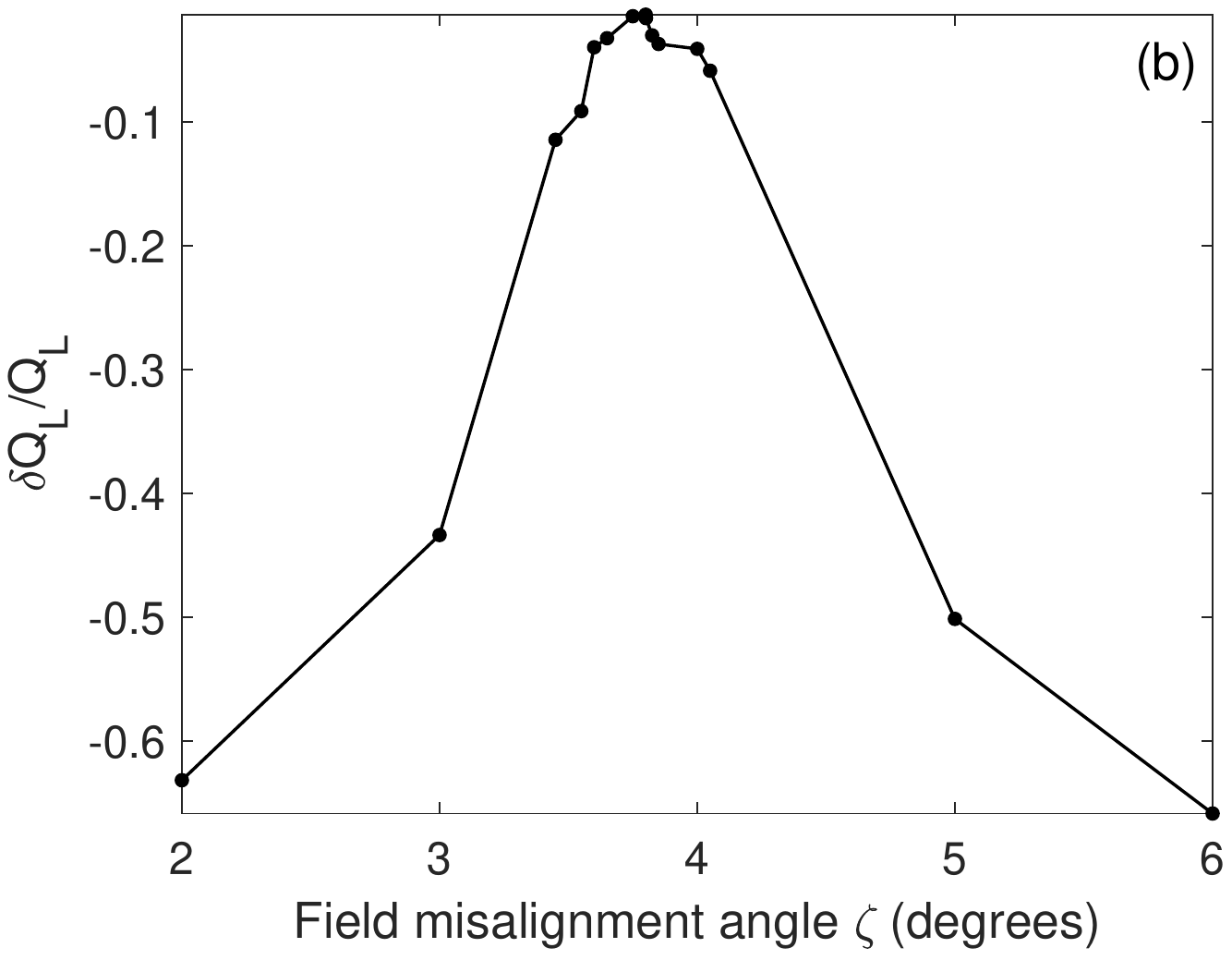}
	\caption{Normalized shifts in (a) resonant frequencies and (b) loaded quality factors after cycling $B$ from $0$ to $60$ mT and back to $0$ as a function of the offset angle $\zeta$. Both $\delta f(\zeta)$ and $\delta Q_L(\zeta)$ are peaked at $\zeta = 3.8^\circ$.}
	\label{fig:rxvsdf0}
\end{figure}

For the NLME measurements, the alignment of the dc field ${\bf B}$ to the plane of the strip is crucial to keep the superconductor in the Meissner state and avoid perpendicular vortices penetrating from the film edges. These vortices caused by the misaligned field reduce the quality factor and give rise to an additional field dependence of $\delta f(B,T)$ unrelated to the NLME. To find the orientation of the magnet which produces ${\bf B}$ parallel to the film plane and the minimum amount of trapped flux, the loaded quality factor $Q_L({\bf B})$ and $\delta f({\bf B})$ were measured as functions of the out-of-plane field angle $\zeta$. We first measured the initial values of $f_{0i}$ and $Q_{Li}$ at zero field, ramped the field up to 60 mT and down to zero, and measured the resulting values of $f_{0a}$ and $Q_{La}$ affected by the number of vortices trapped in the process. The sample was then thermal cycled above $T_c =17.2$ K at zero field to flush out trapped vortices. Measurements were repeated after the magnet was adjusted to a new angle. Shown in Fig. \ref{fig:rxvsdf0} are the normalized shifts $\delta f_0/f_0=(f_{0a}-f_{0i})/f_0$ and $\delta Q_L/Q_L = (Q_{La}-Q_{Li})/Q_L$ as  functions of the magnet angle $\zeta$. Both $\delta f_0(\zeta)/f_0$ and $\delta Q_L(\zeta)/Q_L(0)$ peaked at $\zeta = 3.8^\circ$ which we adopted as a magnet orientation producing the dc field parallel to the plane of the film. This procedure is similar to that which was used in Ref. \onlinecite{ggc}.  

\begin{figure}[ht]
	\includegraphics[trim=0mm 10mm 10mm 8mm, clip, width=1.0\linewidth]{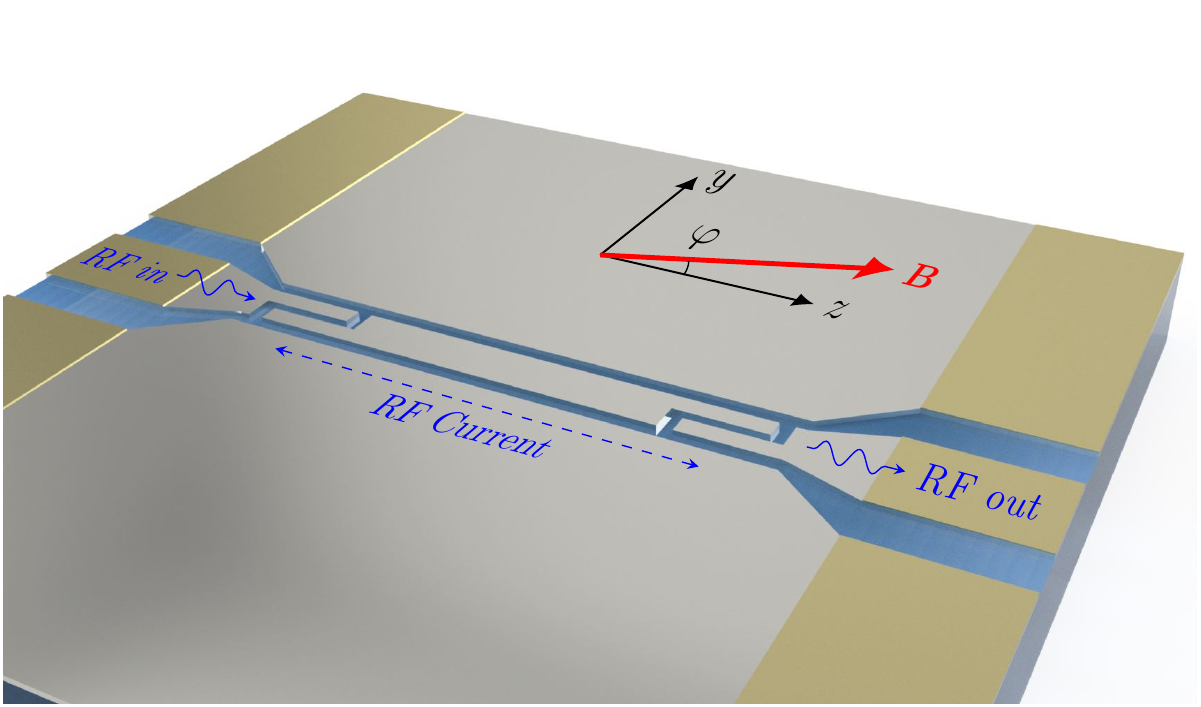}
	\caption{The coordinate system for the magnet orientation with respect to the direction of the rf current on the coplanar resonator.}
	\label{fig:resizenb3snchip2}
\end{figure}

Having aligned the magnet, we measured $f_0(T,B)$ as a function of in-plane dc field up to 200 mT at parallel $\varphi = 0^\circ$ and perpendicular $\varphi = 90^\circ$ field orientations with respect to the strip indicated by Fig. \ref{fig:resizenb3snchip2} at temperatures between 7 K and 12 K. After measurements at a given temperature were completed, the sample was warmed up above $T_c$ to expel any trapped vortices. For each field and temperature point, we repeated the measurement over 50 times, and the average $f_0(B,T)$ was calculated. 

\subsection{Temperature Dependence} 
\begin{figure}[!htb]
	\centering 
	\includegraphics[width=1.0\linewidth]{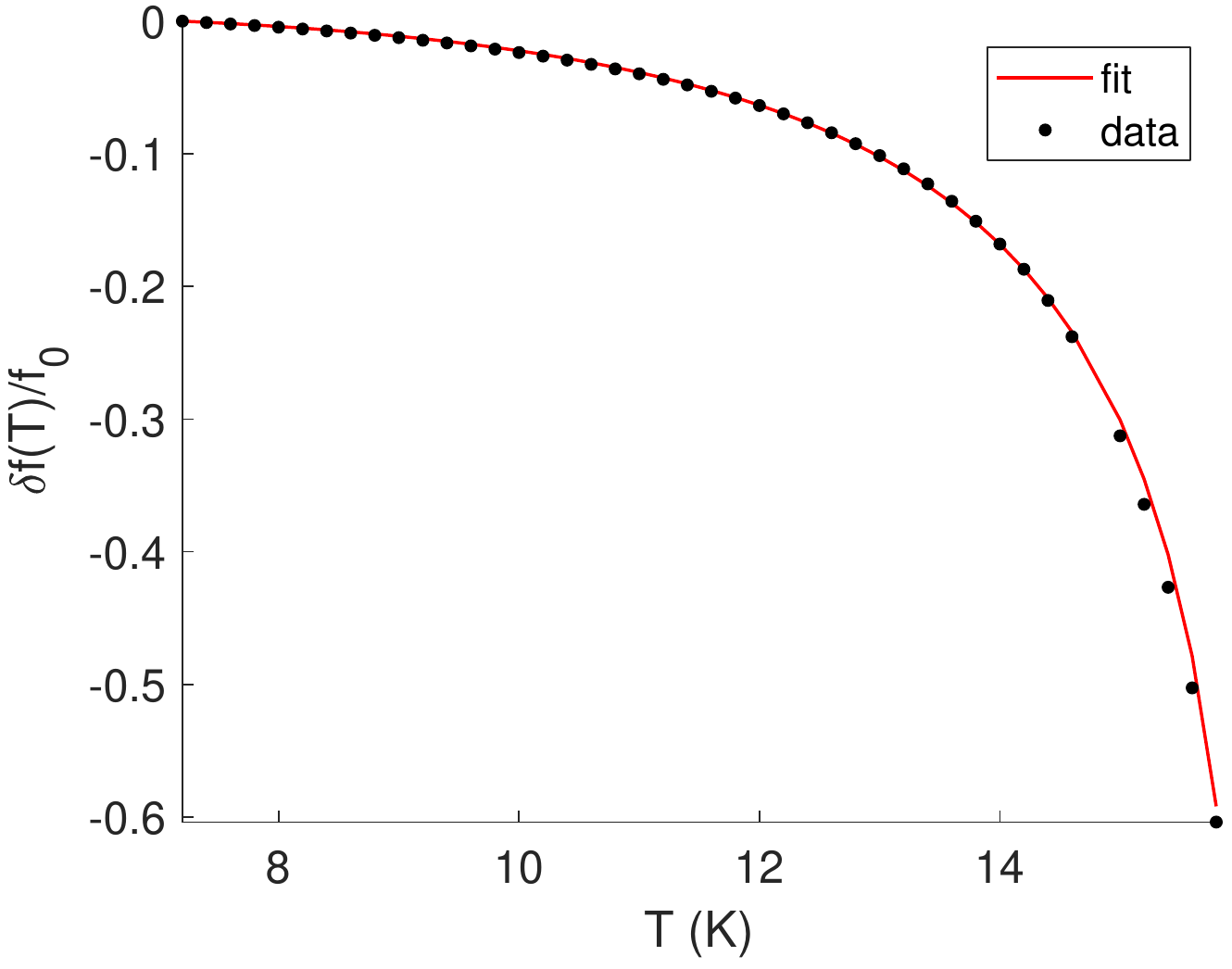} 
	\caption{The normalized temperature-dependent part of the resonant frequency. The fit of the data to Eq. \ref{dffT} gives $\lambda(0) = 353$ nm.} 
	\label{fig:tvsdf} 
\end{figure}

The temperature-dependence of $\lambda(T)$ can be obtained from the measurements of the resonance frequency $f_0(T)$. Using Eqs. (\ref{Lk}) and (\ref{dffT}) we extracted $\lambda(T)$ from the measured $f_0(T)$ by fitting the temperature dependence of a relative frequency shift $\delta f/f_0=[f_0(T)-f_0(7K)]/f_0(7K)$ with the conventional two-fluid approximation of $\lambda(T)=\lambda(0)[1-(T/T_c)^4]^{-1/2}$. Shown in Fig. \ref{fig:tvsdf} is the temperature dependent part of $f_0(T)$ along with the fit with Eq. \ref{dffT}. The fit gives $\lambda(0) = 353$ nm, well above the London penetration depth $\lambda(0) \approx 90$ nm for a clean stoichiometric Nb$_3$Sn ~\cite{hein}. The latter may result from nonstoichimetric inclusions which cause a slight reduction of $T_c$ in our films \cite{chris,ml}. For $\lambda(0)=352$ nm, $w=15\,\mu$m and $d=50$ nm, the kinetic inductance $L_k=\mu_0\lambda^2(T)/wd\simeq 200$ nH/m at $9$ K accounts for about $1/3$ of the total inductance $L=L_g+L_k$ with $L_g=420.5$ nH/m. 

Another factor contributing to the large value of $\lambda(0)$ is the grain boundary structure of our Nb$_3$Sn films shown in Fig. \ref{fig:afm}.  It has been well-established that Sn depletion at GBs ~\cite{gb1,gb2,gb3,gb4} results in weak Josephson coupling of crystalline grains, which has been used to optimize pinning of vortices by GBs in Nb$_3$Sn conductors \cite{arno,pin}. In turn, the weakly-coupled GBs facilitate preferential penetration of the magnetic field along the GB network causing an increase of the global $\lambda$, as is characteristic of many superconductors with short coherence length, including Nb$_3$Sn, cuprates and pnictides ~\cite{grb,gb}.     

\subsection{Field Dependence} 

\begin{figure} [!htb]
	\includegraphics[width=1.0\linewidth]{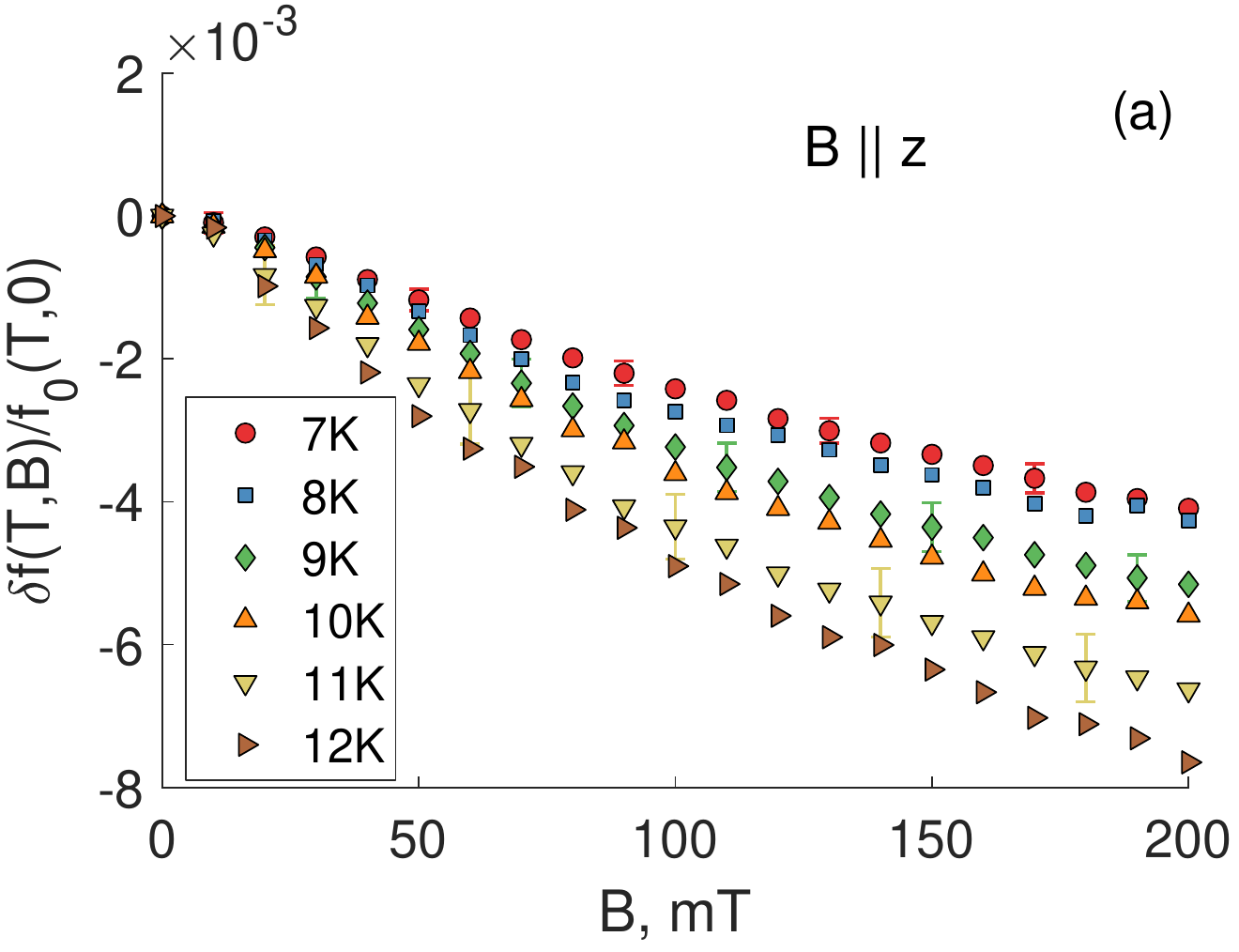} \\ 
	\includegraphics[width=1.0\linewidth]{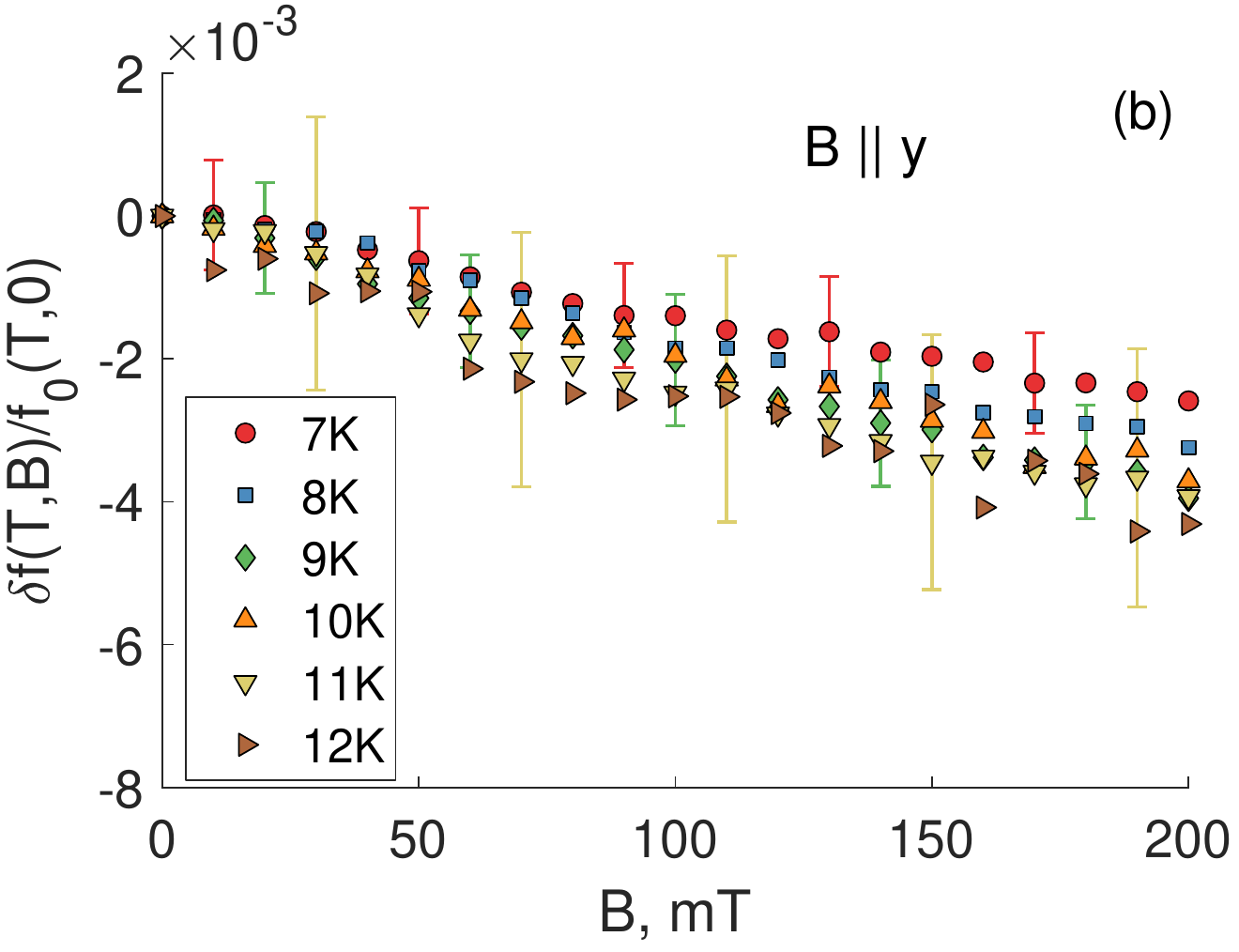} 
	\caption{Normalized frequency shift $\delta f_0(T,B)/f_0(T,0)$ as a function of the in-plane dc magnetic field (a) parallel and (b) perpendicular to the 
	strip.} 
	\label{fig:hvsdff} 
\end{figure} 

Shown in Fig. \ref{fig:hvsdff} are the observed field dependencies of the frequency shifts  for in-plane ${\bf B}$ parallel and perpendicular to the strip. In both cases $\delta f(B)$ decreases nearly linearly with $B$ above $30-40$ mT but flattens at lower fields. Here the slope of $\delta f(B)$ for in-plane ${\bf B}$ parallel  to the strip is about twice of the slope of $\delta f(B)$ for in-plane ${\bf B}$ perpendicular to the strip. In the field range $0<B<200$ mT of our measurements the Nb$_3$Sn film of thickness $50$ nm is in the Meissner state as the parallel field $B$ remains below the nominal lower critical field of a vortex in a thin film. Indeed, an estimate of $B_{c1}=(2\phi_0/\pi d^2)\ln(d/\xi)$ in the London approximation~ \cite{Abrikos} with $d=50$ nm and $\xi=5$  nm \cite{orlando,arno} yields $B_{c1}\simeq 1.17$ tesla exceeding $B_c=0.54$ tesla of Nb$_3$Sn ~\cite{orlando,arno}.  The slope of $\delta f(B)$ in Fig. \ref{fig:hvsdff} increases with increasing temperature, consistent with the temperature dependence of $\lambda(T)$.
  
Our $\delta f(B)$ data exhibit significant scatter, as has also been observed in NLME experiments on  cuprate ~\cite{nme3}.  
The error bars in Fig. \ref{fig:hvsdff}  represent a standard deviation from repeated measurements at each data points, $\delta f(B)$ data for ${\bf B} \| y$ having larger error bars as compared to ${\bf B} \| z$. The main contribution to the error bars comes from vibrations of the sample stage and GSG probes originating from the cryocooler. For ${\bf B}$ perpendicular to the strip, longer probe arms had to be used in the NLME measurements, which increases the amplitude of vibrations.  

\section{Contributions to NLME}

In this section we consider different contributions to the resonant frequency shift $\delta f(T,B)$ caused by the in-plane dc magnetic field $B$.  

\subsection{Meissner current pairbreaking}  
We start with the calculation of the contribution of pairbreaking Meissner currents to $\delta f(H)$ using the TDGL equations for a dirty s-wave superconductor \cite{Kopnin,tdgl}: 
\begin{gather}
\tau_{GL}(1+4\tau_E^2\Delta^2)^{-1/2}\left(\frac{\partial}{\partial t}+2ie\Phi+2\tau_E^2\frac{\partial \Delta^2}{\partial t}\right)\Psi \nonumber\\ =\left(1-\frac{\Delta^2}{\Delta_0^2}\right)\Psi+\xi^2\left(\mathbf{\nabla}-2ie\mathbf{A}\right)^2\Psi,
\label{gtdgl}\\
\mathbf{J}=-\frac{\pi\sigma_0}{4eT_c}\Delta^2\mathbf{Q}-\sigma_0\left(\mathbf{\nabla}\Phi+\frac{\partial \mathbf{A}}{\partial t}\right).
\label{j}
\end{gather}
Here $\tau_{GL}=\pi\hbar/8k_B(T_c-T)$, $\xi=[\pi \hbar D/8k_B(T_c-T)]^{1/2}$ is the coherence length, $D$ the electron diffusivity, $\Phi$ is a scalar potential, $\tau_E$ is an energy relaxation time due to electron-phonon scattering \cite{Kopnin}, 
$\Delta_0^2=8\pi^2k_B^2T_c(T_c-T)/7\zeta(3)$, $\sigma_0=2e^2DN(0)$ is the normal state conductivity, $N(0)$ is the density of states at the Fermi surface, and $-e$ is the electron charge. Equations (\ref{gtdgl}) and (\ref{j}) were derived from the kinetic BCS  theory assuming that $\mathbf{Q}(\mathbf{r},t)$ and $\Delta(\mathbf{r},t)$ vary slowly over $\xi_0\simeq (\hbar D/k_BT_c)^{1/2}$, the diffusion length $L_E=(D\tau_E)^{1/2}$ and $\tau_E$ ~\cite{tdgl,Kopnin}, where   
\begin{equation}
\tau_E=\frac{8\hbar}{7\pi\zeta(3)\gamma k_B T_F}\left(\frac{c_s}{v_F}\right)^2\left(\frac{T_F}{T}\right)^3.
\label{tau}
\end{equation}
Here $c_s$ is the speed of longitudinal sound, $v_F$ and $T_F=\epsilon_F/k_B$ are the Fermi velocity and temperature, respectively, and $\gamma$ is a dimensionless electron-phonon coupling constant. For  $c_s/v_F\simeq 10^{-3}$, $T_F\sim 10^5$ K, $T_c=17$ K and $\gamma\simeq 1.5$ ~\cite{orlando}, Eq. (\ref{tau}) yields $\tau_E(T_c)\sim 10$ ps. 
 
For a wide film in a parallel magnetic field, $Q(x)$ and $\Delta(x)$ depend only on the coordinate $x$ across the film, and the TDGL equations in the gauge $\Phi=0$ can be written in the dimensionless form: 
\begin{gather}
(1+g^2\psi^2)^{1/2}\dot{\psi}=(1-q^2)\psi +\psi''-\psi^3,
\label{glq}\\
j = -u\psi^2q-\dot{q},
\label{glj}
\end{gather} 
where $\psi=\Delta/\Delta_0$, $q=Q\xi$, $g=2\Delta_0\tau_E/\hbar$, $j=J/J_0$, $t$ is in units of $\tau_{GL}$, $J_0=\sigma_0/2e\xi\tau_{GL}$, $x$ is in units of $\xi$, the prime and overdot denotes differentiation with respect to $x$ and $t$, respectively, and  $u=\pi^4/14\zeta(3)\approx 5.79$.

For a coplanar resonator of thickness $d<\lambda$ and width $w\gg \lambda$ in a parallel dc field ${\bf B}$ inclined by the angle $\varphi$ to the $z$-axis along the strip, we have:
\begin{equation}
q_{z}=-hx\sin\varphi+a_\omega e^{i\omega t},\qquad q_{y}=hx\cos\varphi,
\label{q} 
\end{equation} 
where $h=B/B_{c2}$, $B_{c2}=\phi_0/2\pi\xi^2$, $a_\omega=A_\omega e^{i\omega t}/A_0$, $A_\omega$ is a small rf vector potential excited along the strip, $A_0=\phi_0/2\pi\xi$, $x=0$ is taken in the middle of the film and the London screening at $d\ll\lambda$ is disregarded.

At $a_\omega=0$ the dc field causes a reduction of $\psi(x)=1-\psi_1(x)$. The equation for a field-induced correction $\psi_1$ is obtained from Eq. (\ref{glq}) in the first order in $h^2\ll 1$: 
\begin{equation}
\psi_1''-2\psi_1=-h^2x^2,\qquad \psi_1'(\pm d/2)=0.
\label{pp1}
\end{equation}
where $d$ is in units of $\xi$. The squared order parameter averaged over the film thickness, $\bar{\psi}^2 \simeq 1-2d^{-1}\int_{-d/2}^{d/2}\psi_1dx$, is obtained by integrating Eq. (\ref{pp1}) over $x$:
\begin{equation}
\bar{\psi}^2=1-\frac{d^2h^2}{12}.
\label{barp}
\end{equation}
To calculate a linear response current $\delta I_\omega e^{i\omega t}$ induced by a weak $a_\omega e^{i\omega t}$ in the presence of a parallel dc field, we linearize Eq. (\ref{glj}) in $a_\omega$:
\begin{equation}
\frac{\delta I_\omega}{wJ_0} = -ud\bar{\psi}^2 a_\omega+2hu\sin\varphi\!\int_{-d/2}^{d/2}\!\!x\delta\psi(x)dx -id\omega a_\omega.
\label{dI}
\end{equation}
This shows that $\delta\psi$ is coupled with the dc Meissner current flowing along the strip due to the field component $B_y=B\sin\varphi$. Here $\delta\psi(x)\propto a_\omega $ satisfies the following equation obtained from Eq. (\ref{glq})  
linearized with respect to small $\delta\psi$ and $a_\omega$:
\begin{gather}
\delta\psi''-k_\omega^2\delta\psi=-2a_\omega hx\sin\varphi,\qquad \delta\psi'(\pm d/2)=0,
\label{ps1}\\
k_\omega^2=2+i\omega\tau,\qquad \tau=\tau_{GL}\sqrt{1+(2\tau_E\Delta/\hbar)^2}.
\label{k}
\end{gather}
The solution of Eq. (\ref{ps1}) is given by:
\begin{gather}
\delta\psi(x)=\sum_{n=0}^\infty A_n\sin q_nx,\qquad q_n=\frac{\pi}{d}(2n+1),
\label{p1}\\
A_n=\frac{8(-1)^n a_\omega h\sin\varphi}{dq_n^2(k_\omega^2+q_n^2)}. 
\label{An} 
\end{gather}
Inserting $\delta\psi(x)$ in Eq. (\ref{dI}) and integrating over $x$ yields:
\begin{equation}
\frac{\delta I_\omega}{dwJ_0} =-\biggl[u\bar{\psi}^2-\sum_{n=0}^{\infty}\frac{32ud^{2}h^2\sin^2\varphi}{\pi^{4}(2n+1)^{4}(2+i\omega r+q_{n}^{2})}+i\omega\biggr]a_\omega
\label{resp}
\end{equation}
The sum in Eq. (\ref{resp}) converges rapidly, so $q_n^2 \propto (\xi/d)^2\ll 1$ in the denominator can be neglected in films with $\xi\ll d\ll\lambda$. Using $\sum_{n=0}^\infty (2n+1)^{-4}=\pi^4/96$ and restoring the original units, we obtain the linear response current $\delta I_\omega$ induced by the ac vector potential $A_\omega$: 
\begin{equation}
\delta I_\omega =-\left( \frac{1}{\mu_0\tilde{\lambda}_\omega^2}+i\omega \sigma_0\right)dwA_\omega,
\label{re}
\end{equation}  
where the complex London penetration depth is given by: 
\begin{gather}
\frac{1}{\tilde{\lambda}_{\omega}^{2}}=\frac{1}{\lambda^{2}}\bigg[1-\frac{1}{3}\left(\frac{2\pi\xi dB}{\phi_{0}}\right)^{2} 
\left(\frac{1}{4}+\frac{\sin^{2}\varphi}{2+i\omega\tau}\right)\bigg].
\label{lame} 
\end{gather}
Here the NLME term $\propto B^2$ depends on $\omega$ and the field orientation angle $\varphi$.  The maximum NLME contribution to 
$\lambda_\omega$ occurs at $\varphi=\pi/2$ when ${\bf B}\perp z$ and the rf current is parallel to the dc Meissner current. At $\omega\tau\ll 1$ the angular 
dependence of $\lambda_\omega(\varphi)$ in Eq. (\ref{lame}) reduces to that was obtained previously in a quasi-static limit ~\cite{ggc}. 

The imaginary part of $\tilde{\lambda}_\omega^2$ contributes to the dynamic conductivity $\sigma_\omega$. Denoting $\lambda_\omega^2=\mbox{Re}\tilde{\lambda}_\omega^2$, and separating real and imaginary parts in Eqs. (\ref{re}) and (\ref{lame}), yields:
\begin{gather}
\lambda_\omega^2=\biggl[1+\frac{1}{3}\left(\frac{2\pi\xi dB}{\phi_{0}}\right)^2\left(\frac{1}{4}+\frac{2\sin^{2}\varphi}{4+\omega^2\tau^2}\right)\biggr]\lambda^2,
\label{lamb} \\
\sigma_\omega=\sigma_0+\frac{1}{3\mu_0\lambda^2}\left(\frac{2\pi\xi dB}{\phi_{0}}\right)^2\frac{\tau\sin^{2}\varphi}{4+\omega^2\tau^2}.
\label{sig}
\end{gather}
The NLME field correction to the kinetic inductance $L_k^M=\mu_0\lambda^2(B)/dw$ is given by:
\begin{equation}
\delta L_k^M= \frac{\mu_0\lambda^2}{3 dw}\left(\frac{\pi\xi dB}{\phi_{0}}\right)^2\left[1+\frac{2\sin^{2}\varphi}{1+(\omega\tau/2)^2}\right].
\label{lkd}
\end{equation}
As follows from Eqs. (\ref{sig}) and (\ref{lkd}), the NLME correction to $\sigma_\omega$ remains finite at $T_c$, while $\delta L_k^M\propto (T_c-T)^{-2}$ increases stronger than the zero-field kinetic inductance $L_k^M=\lambda^2/d\propto (T_c-T)^{-1}$ as $T\to T_c$. The dependencies of $\lambda_\omega$ and $\sigma_\omega$ on the orientation of ${\bf B}$ persist as long as $\omega\tau\lesssim 1$ and disappear at $\omega\tau\gg 1$. The latter occurs both at $T\to T_c$ where $\tau_{GL}(T)$ diverges and at low temperatures where $\tau_E(T)\propto T^{-3}$ increases strongly. 

We estimate $\delta\lambda(B)=\lambda(B)-\lambda=(\pi\xi dB/\phi_0)^2/2$ at $\varphi=\pi/2$ and $\omega\tau\ll 1$ for $d=50$ nm and $\xi=5$ nm. Here $\delta\lambda/\lambda= 7.7\times 10^{-4}$ at $B=100$ mT, which translates to $\delta\lambda=0.27$ nm at $\lambda=350$ nm.
If the total $L$ is dominated by the kinetic inductance, Eq. (\ref{lkd}) yields the maximum NLME frequency shift:
\begin{equation}
\frac{\delta f}{f}=-\frac{1}{6}\left(\frac{\pi\xi dB}{\phi_{0}}\right)^2\left[1+\frac{2\sin^{2}\varphi}{1+(\omega\tau/2)^2}\right].
\label{dfm}
\end{equation}
In superconductors with $\kappa\gg 1$ the maximum $\delta f(B_{c1})/f_0$ in a thin film is much greater than the maximum NLME bulk shift $\delta \lambda/\lambda=(B_{c1}/B_c)^2/8=(\ln\kappa/4\kappa)^2$ which follows from Eq. (\ref{gl}). For  $d=50$ nm $\xi=5$ nm, $\varphi=\pi/2$ and $B=100$ mT, we obtain $\delta f(B)/f_0=(\pi\xi dB/\phi_0)^2/2\simeq 8\times 10^{-4}$, well below the observed $\delta f/f_0$. Taking the geometrical inductance into account further reduces $\delta f/f_0$ by a factor $\simeq 3$. Moreover, according to Eq. (\ref{lamb}) the field-induced shift $\delta \lambda$ at $\varphi=\pi/2$ for which the rf currents are parallel to the dc Meissner currents is 3 times larger than $\delta \lambda$ at $\varphi=0$. This is inconsistent with the experimental data shown in Fig. \ref{fig:hvsdff} where the slope of $\delta f(B)/f_0$ at $\varphi=0$ is about 2 times larger than for $\delta f(B)/f_0$ at $\varphi=\pi/2$. Thus, not only is the Meissner pairbreaking too weak to account for the observed $\delta f/f_0$ but it yields the field and orientational dependencies of $\delta f({\bf B})/f_0$ inconsistent with our experimental data on Nb$_3$Sn. 
\subsection{Grain boundary contribution}

A significant contribution to $L_k$ can come from local non-stoichiometry, strains, and grain boundaries (GBs) in  polycrystalline Nb$_3$Sn. Particularly, the well-known Sn depletion at GBs ~\cite{gb1,gb2,gb3,gb4} results in weak Josephson coupling of grains in Nb$_3$Sn. Our polycrystalline films have lateral grain sizes $l_2 \sim 0.1-1~\mu$m (see Fig. \ref{fig:afm}) and local nonstoichiometry causing inhomogeneities of superconducting properties on the same length scales~ \cite{chris,ml}. If weakly coupled GBs are regarded as planar Josephson junctions (JJs), each GB has a kinetic inductance ~\cite{BP}
\begin{equation}
L_J=\frac{\phi_0}{2\pi I_c\cos\theta},
\label{kinj}
\end{equation}
where $I_c(T)$ is a critical current of the JJ. The field dependence of $L_J(B)$ is determined by the phase difference $\theta({\bf r})$ induced 
by the dc field on a GB.  Because GBs in Nb$_3$Sn can have broad distributions of sizes and $I_c$ values ~\cite{gb}, low-$I_c$ GBs can significantly increase $L_J$ if they form interfaces blocking the cross-section of the film.

An array of weakly-coupled GBs can be modeled by a Hamiltonian of a granular superconductor ~\cite{net,stroud1,stroud2}
\begin{equation}
{\cal H}=-\sum_{ij} J_{ij}\cos(\chi_i-\chi_j-A_{ij}),
\label{ham}
\end{equation} 
where the coupling energies $J_{ij}=\hbar I_c^{ij}/2e$ of the $i$-th and $j$-th grains are proportional to 
the respective intergrain Josephson critical currents $I_c^{ij}$, $\chi_i$ and $\chi_j$ are phases of the 
superconducting order parameters in the grains, and  the magnetic phase factors $A_{ij}$ are given by:
\begin{equation}
A_{ij}=\frac{2\pi}{\phi_0}\int_i^j{\bf A}\cdot d{\bf l}.
\label{Aij}
\end{equation}  
Monte-Carlo simulations of the Hamiltonian (\ref{ham}) of a disordered XY model have shown that the helicity moduli and the global kinetic inductance 
change nearly linearly with the dc magnetic field at $B\lesssim \phi_0/l_2^2 $ and exhibit strong fluctuations as functions of $B$ 
due to transitions between many metastable states in finite JJ arrays ~\cite{stroud1,stroud2}. These results appear qualitatively consistent with the observed field dependence of $\delta f(B)/f_0$ shown in Fig. \ref{fig:hvsdff}.

To get an insight into the nearly linear decrease of $\delta f(B)$ with $B$ shown in Fig. \ref{fig:hvsdff}, we use a mean-field model 
in which $\cos(\chi_i-\chi_j-A_{ij})$ is replaced with an averaged value $\langle\cos\theta\rangle$. Then a disturbance 
$\delta\theta$ induced by a weak rf current $J_\omega e^{i\omega t}$ on an overdamped GB is described by the dynamic equation of the 
RSJ model
\begin{equation}
\tau_J\delta\dot{\theta} +\langle\cos\theta\rangle\delta\theta=(J_\omega/J_c) e^{i\omega t},
\label{rsj}
\end{equation}
where $\tau_J$ is a relaxation time caused by ohmic quasiparticle current through JJ, and $\langle ...\rangle$ denotes averaging over orientations of the GB planes:
\begin{equation}
\langle \cos\theta \rangle=\frac{1}{NS}\sum_s n_{zs}^2\int_S \cos\theta_s ({\bf r})dS
\label{av}
\end{equation} 
Here the summation goes over all GBs, and $\theta_s({\bf r})$ on a planar GB smaller than the Josephson length $\lambda_J=(\phi_0/2\pi\mu_0 d J_c)^{1/2}$ is determined by ~\cite{BP}
\begin{equation}
\nabla\theta=\frac{2\pi \Lambda}{\phi_0}[{\bf B}\times {\bf n}],
\label{grad}
\end{equation}
where ${\bf n}$ is a unit vector normal to the GB surface, and $\Lambda\sim d$ for a JJ in a thin film $(d\lesssim \lambda)$ in a parallel magnetic field ~\cite{BP,jj1,jj2}. The factor 
$n_z^2$ in Eq. (\ref{av}) takes into accounts that only the perpendicular component of the rf current $n_zJ_\omega=J_c\delta\theta$
causes the Josephson voltage $V_\omega\hbar\delta\dot{\theta}n_z/2e$ along $J_\omega$. 
From Eq. (\ref{rsj}), we obtain $\delta\theta=J_\omega e^{i\omega t}/(\langle\cos\theta\rangle+i\omega\tau_J)J_c$ and the impedance $Z_J=NV_\omega/I_\omega$ of $N$ grain boundaries:
\begin{equation}
Z=\frac{i\omega NR\langle L_J\rangle}{R+i\omega\langle L_J\rangle},\qquad \langle L_J\rangle=\frac{\phi_0}{2\pi I_c\langle\cos\theta\rangle}
\label{imp}
\end{equation}  

\begin{figure}[!htb]
	\centering
    \includegraphics[scale=0.4,trim={80mm 55mm 55mm 40mm},clip,width=1.0\linewidth]{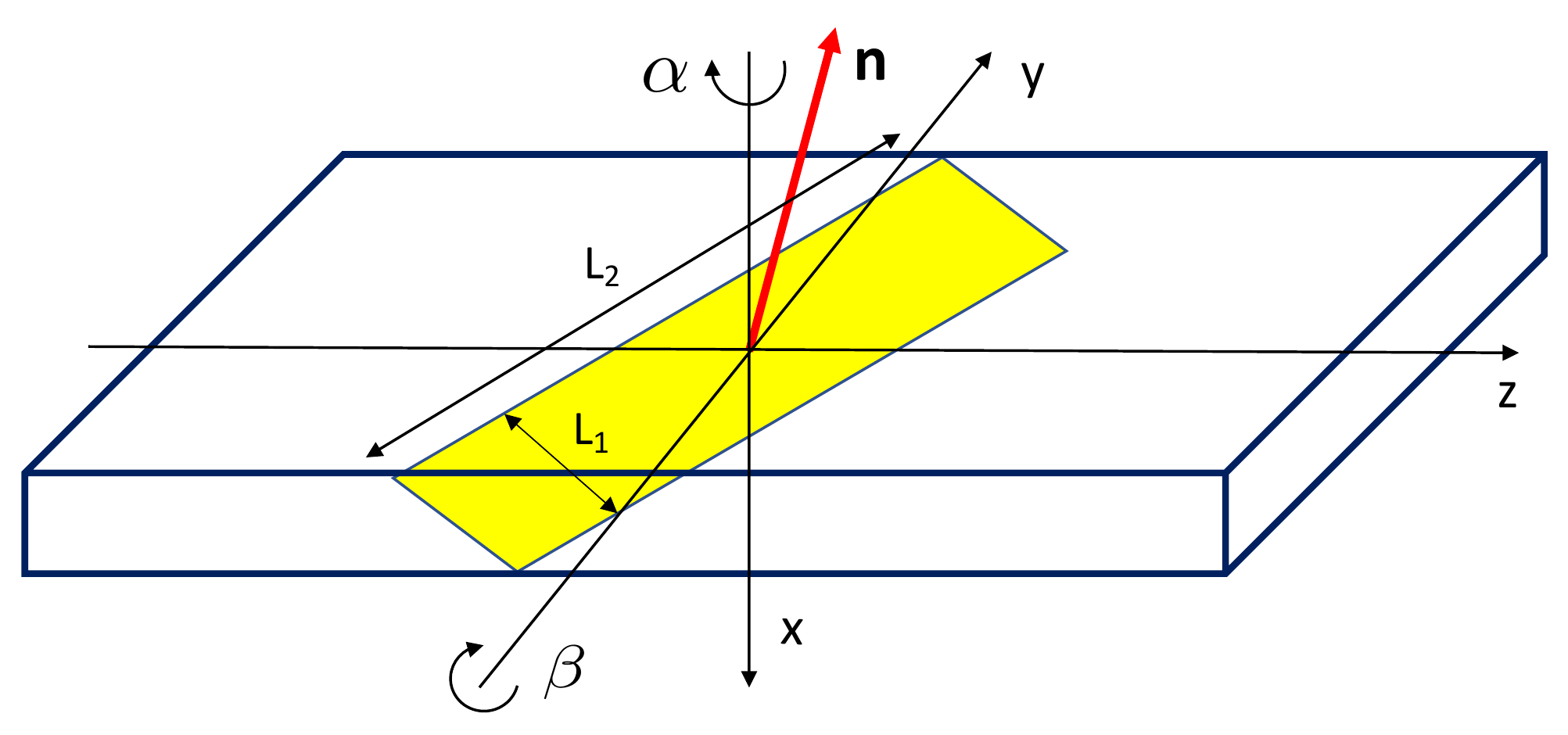}
	\caption{Geometry of a rectangular tilted GB (yellow). The red arrow shows the normal to the GB plane.} 
	\label{Fig.X}
\end{figure}

We calculate $\langle\cos\theta\rangle$ for randomly-oriented planar GBs parameterized by the Euler angles $\alpha$ and $\beta$ shown in Fig. \ref{Fig.X}. As shown in Appendix \ref{B}, averaging $\cos\theta$ over the area $S$ of a tilted GB in a magnetic field gives:
\begin{gather}
\bar{c} =\frac{1}{S}\int\cos\theta dS=
\frac{\sin(q_{1}l_{1})\sin(q_{2}l_{2})}{q_{1}l_{1}q_{2}l_{2}},
\label{barc}
\end{gather}
where
\begin{gather}
q_{1}=\frac{\pi B}{\phi_{0}}\Lambda\sin\alpha,\quad q_{2}=\frac{\pi B}{\phi_{0}}\Lambda\sin\beta\cos\alpha,\quad{\bf B}\|z
\label{qz}\\
q_{1}=\frac{\pi B}{\phi_{0}}\Lambda\cos\alpha,\quad q_{2}=\frac{\pi B}{\phi_{0}}\Lambda\sin\beta\sin\alpha,\quad{\bf B}\|y
\label{qy}
\end{gather}
There is a significant difference of GB lengths $l_1$ and $l_2$ in our Nb$_3$Sn coplanar resonator with $d\ll w$. Here $l_1$ is smaller or of the order of the film thickness, $l_1\lesssim d\simeq 50$ nm, whereas lateral GB lengths $l_2$ are in a submicron range $l_2\sim 0.1-1\,\mu$m (see Fig. \ref{fig:afm}) so that $l_1\sim (10^{-2}-10^{-1})l_2$.  As a result, $\sin(q_1l_1)/l_1q_1$ in Eq. (\ref{barc}) remains close to $1$ in the field region $B\lesssim \phi_0/\pi\Lambda l_1\sim \phi_0/\pi d^2\sim B_{c1}$ of our measurements, while $\sin(q_2l_2)/l_2q_2$ has a strong field dependence at $B> B_0\simeq \phi_0/\pi d l_2$. Here $B_0\simeq 25$ mT at $l_2=0.5\,\mu$m and $d=50$ nm. As a result, Eq. (\ref{barc}) at $B<B_{c1}$ simplify to
\begin{gather}
\bar{c}=\frac{\sin(b\cos\alpha\sin\beta)}{b\cos\alpha\sin\beta}, \qquad{\bf B}\|z
\label{cz} \\
\bar{c}=\frac{\sin(b\sin\alpha\sin\beta)}{b\sin\alpha\sin\beta}, \qquad{\bf B}\|y
\label{cy}\\
b=B/B_0,\qquad B_0=\phi_0/\pi\Lambda l_2.
\end{gather}
Here only tilted GBs which are not perpendicular to the film plane $(\beta\neq 0)$ contribute to the strong field dependence of $\bar{c}(B)$. The average $\langle\cos\theta\rangle$ over all GB orientations in Eq. (\ref{av}) can be written in the form 
\begin{equation}
\langle\cos\theta\rangle=\sum_{i,j}P_{ij}\bar{c}_{ij}\cos^2\alpha_i\cos^2\beta_j,
\label{avv}
\end{equation}
where $\bar{c}_{ij}$ is given by either Eq. (\ref{cz}) or (\ref{cy}), depending on the direction of ${\bf B}$, $i$ and $j$ label different GBs, $n_z^{ij}=\cos\alpha_i\cos\beta_j$, and  $P_{ij}$ is a 
probability distribution of $\alpha_i$ and $\beta_j$ normalized by $\sum_{ij}P_{ij}=1$.

\begin{figure}[!htb]
	\centering
	\includegraphics[scale=0.5,trim={33mm 70mm 20mm 75mm},clip]{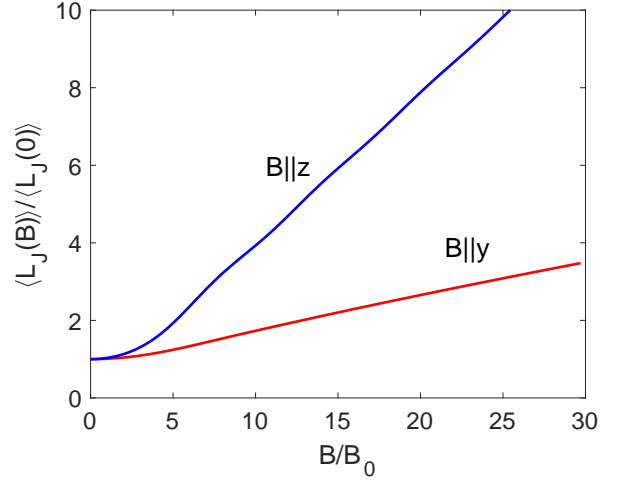}
	\caption{The field-dependencies of the kinetic inductance $\langle L_J(B)\rangle$ calculated from Eqs. (\ref{imp}), (\ref{cosz}) and 
	(\ref{cosy}) for random orientation of GB planes, and the dc magnetic field applied parallel $({\bf B}\|z)$ and perpendicular $({\bf B}\|y)$ 
	to the strip line. } 
	\label{Fig.Y}
\end{figure}

The angular distributions of GBs is affected by the crystalline texturing during the 
film growth and other materials factors. We consider here a simple case of 
randomly-oriented GBs with equal probabilities of all $\alpha_i$ and $\beta_j$. Then  Eq. (\ref{avv}) reduces to the integrals 
(\ref{cosz}) and (\ref{cosy}) given in Appendix \ref{B}. Numerical calculation of these integrals for  
$\langle\cos\theta\rangle$ yields the field dependencies of $\langle L_J\rangle$ shown in Fig. \ref{Fig.Y}.   
  
As follows from Fig. \ref{Fig.Y}, grain boundaries can radically change the field dependence of the kinetic inductance as compared to the NLME caused by the Meissner pairbreaking. First, the GB contribution $\langle L_J(B)\rangle$ is quadratic in $B$ only at very low fields $B\lesssim B_0\ll B_c$ and exhibits a nearly linear field dependence at $B \gtrsim B_0\ll B_c$, whereas the Meissner pairbreaking gives $\delta L_k\propto B^2$ all the way to $B=B_{c1}$. Second, the field ${\bf B}\| z$ applied along the strip causes stronger increase of $\langle L_J(B)\rangle$ than the transverse field ${\bf B}\| y$. This is the opposite of the orientational field dependence of $\delta L_k^M(B)$ described by Eq. (\ref{lkd}) and observed on Nb coplanar resonator ~\cite{ggc}. Yet both features of $\langle L_J(B)\rangle$ are in agreement with our experimental data on polycrystalline Nb$_3$Sn shown in Fig. \ref{fig:hvsdff}.

\section{Discussion}

To compare the contributions of Meissner pairbreaking and weakly-coupled GBs to $\delta f(B)/f_0$,  we evaluate the GB kinetic inductance $L_k^J$ per unit length of the strip in the above mean-field model. If GBs have the same critical current density $J_c$, dimensions $d\times l_2$ but different orientations, $L_k^J\sim \langle L_J\rangle/l_2$, where  $\langle L_J\rangle$ is given by Eq. (\ref{kinj}) with $I_c\sim dwJ_c$. As a result,
\begin{equation}
L_k^J\sim \frac{\phi_0}{2\pi J_c wdl_2\langle \cos\theta\rangle}\sim L_k^M\times \frac{\xi J_d}{l_2J_c}
\label{rat}
\end{equation}
Here $J_d=\phi_0/2\pi\mu_0\lambda^2\xi$ is of the order of the GL depairing current density. The kinetic inductance is dominated by weakly-coupled GBs if $J_c\ll \xi J_d/l_2$. For our Nb$_3$Sn films with $\xi\simeq 5$ nm and $l_2\simeq 200$ nm, the GB contribution dominates if $J_c\lesssim 10^{-2}J_d$.

The effective penetration depth $\lambda$ extracted from the measured kinetic inductance is affected by GBs.
The Meissner contribution $L_k^M$ is determined by the London penetration depth $\lambda$ in the dirty limit~\cite{Kopnin}:
\begin{equation}
L_k^M=\frac{\mu_0\lambda^2}{dw}=\frac{\hbar\rho_s}{\pi dw\Delta(T)}\coth\frac{\Delta(T)}{2 k_B T},
\label{lam0}
\end{equation}
where $\rho_s$ is the normal state resistivity. If GBs can be modeled as S-I-S Josephson junctions~ \cite{BP}, their contribution to $L_k$ can be evaluated from Eq. (\ref{rat}):
\begin{equation}
L_k^J\sim \frac{\hbar}{eJ_c wdl_2}\simeq \frac{\hbar R_\perp}{dwl_2\Delta(T)}\coth\frac{\Delta(T)}{2 k_B T},
\label{lamj}
\end{equation}
where $R_\perp$ is the tunneling resistance of the JJ per unit area. Defining the effective penetration depth $\tilde{\lambda}$ in the total kinetic inductance $L_k=L_k^M+L_k^J=\tilde{\lambda}^2/dw$ and combining Eqs. (\ref{lam0}) and (\ref{lamj}), we obtain:
\begin{equation}
\tilde{\lambda}^2(T)=\frac{\hbar}{\pi\Delta(T)}\left[\rho_s+C_1\frac{R_\perp}{l_2}\right]\coth\frac{\Delta(T)}{2 k_B T},
\label{lamm}
\end{equation}
where the factor $C_1\sim 1$ accounts for details of the shape and angular distributions of GBs. In the S-I-S model the temperature dependencies of the 
GB and Meissner contributions to $\tilde{\lambda}$ are the same. This is no longer the case if GBs are proximity-coupled S-N-S junctions ~\cite{BP} for which $J_c\propto (1-T/T_c)^2$. The S-N-S scenario may be more relevant for Nb$_3$Sn in which strongly-coupled GBs can transmit high current densities which are still well below the depairing limit ~\cite{arno,gb1,gb2,gb3,gb4,pin}.  

The above estimate pertains to $B=0$ but the field-induced frequency shift $\delta f/f_0$ is determined by small field - dependent corrections $\delta L_k^M$ and $\delta L_k^J$. Here $\delta L_k^M$ is given by Eq. (\ref{lkd}) and $\delta L_k^J\sim (\mu_0\lambda^2/dw)(\xi J_d/l_2J_c)(B/B_0)$ at $B\gtrsim B_0\simeq \phi_0/3\pi dl_2$, as shown in Fig. \ref{Fig.Y}. Hence, 
\begin{equation}
\frac{\delta L_k^M}{\delta L_l^J}\sim \frac{B\xi  dJ_c}{\phi_0J_d},\qquad B\gtrsim B_0
\label{ration}
\end{equation}
This ratio is independent of the lateral GB sizes and is much smaller than 1 since $J_c\ll J_d$, and $Bd\xi/\phi_0<2\times 10^{-2}$ at $B<200$ mT, 
$\xi=5$ nm and $d=50$ nm. Thus, the NLME field-dependent frequency shift in our polycrystalline Nb$_3$Sn coplanar resonators is dominated by grain boundaries, even if their contribution to the total kinetic inductance at $B=0$ is much smaller than that of the Meissner currents. This conclusion is consistent not only with the observed nearly linear field dependence of $\delta f/f_0$ but also with the fact that the slope of $\delta f/f_0$ at ${\bf B}$ along the strip is larger than the slope of $\delta f/f_0$ at ${\bf B}$ perpendicular to the strip.  

\begin{figure} [ht]
	\includegraphics[scale=0.45,trim={25mm 70mm 0mm 60mm},clip]{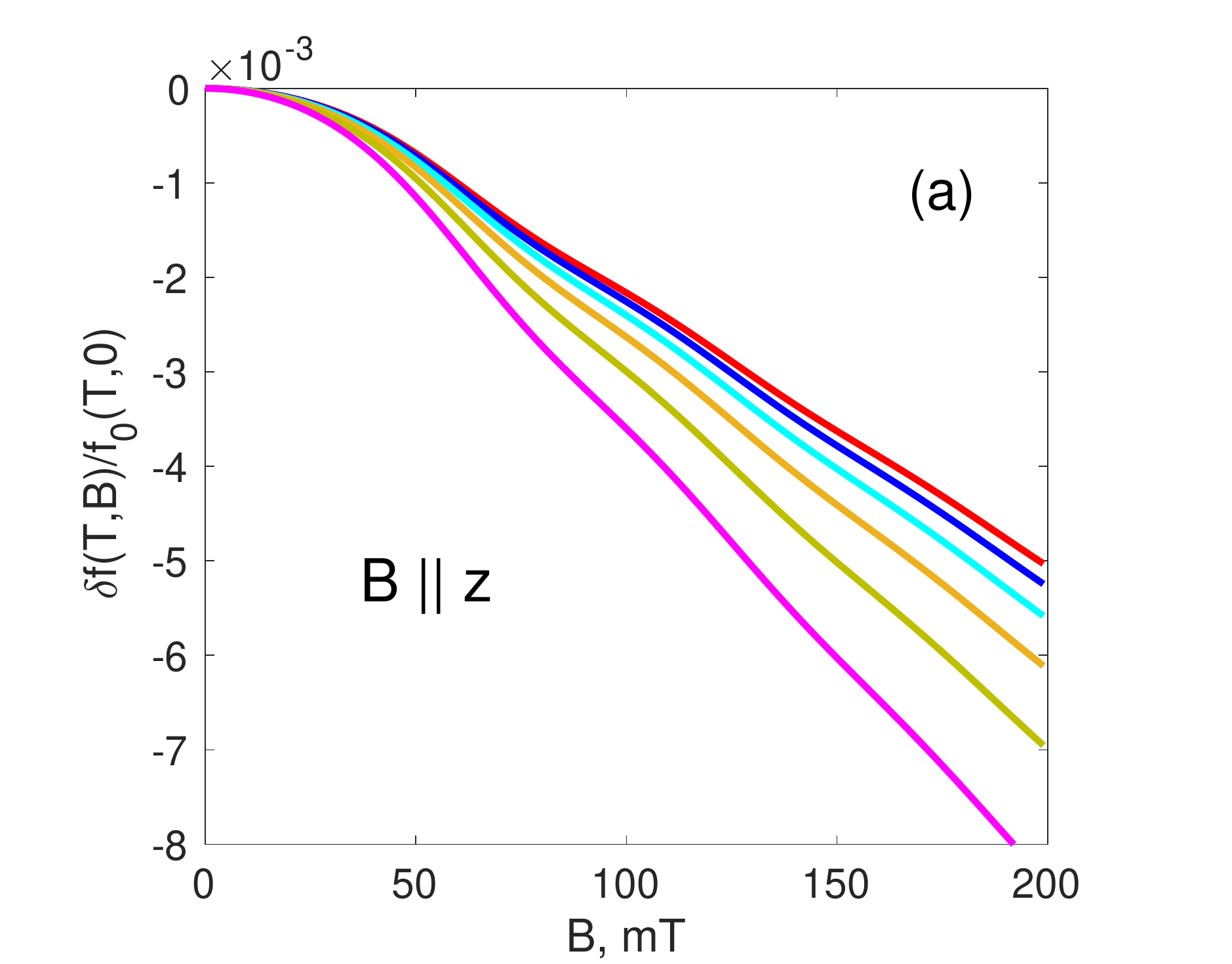} \\ 
	\includegraphics[scale=0.45,trim={25mm 70mm 0mm 60mm},clip]{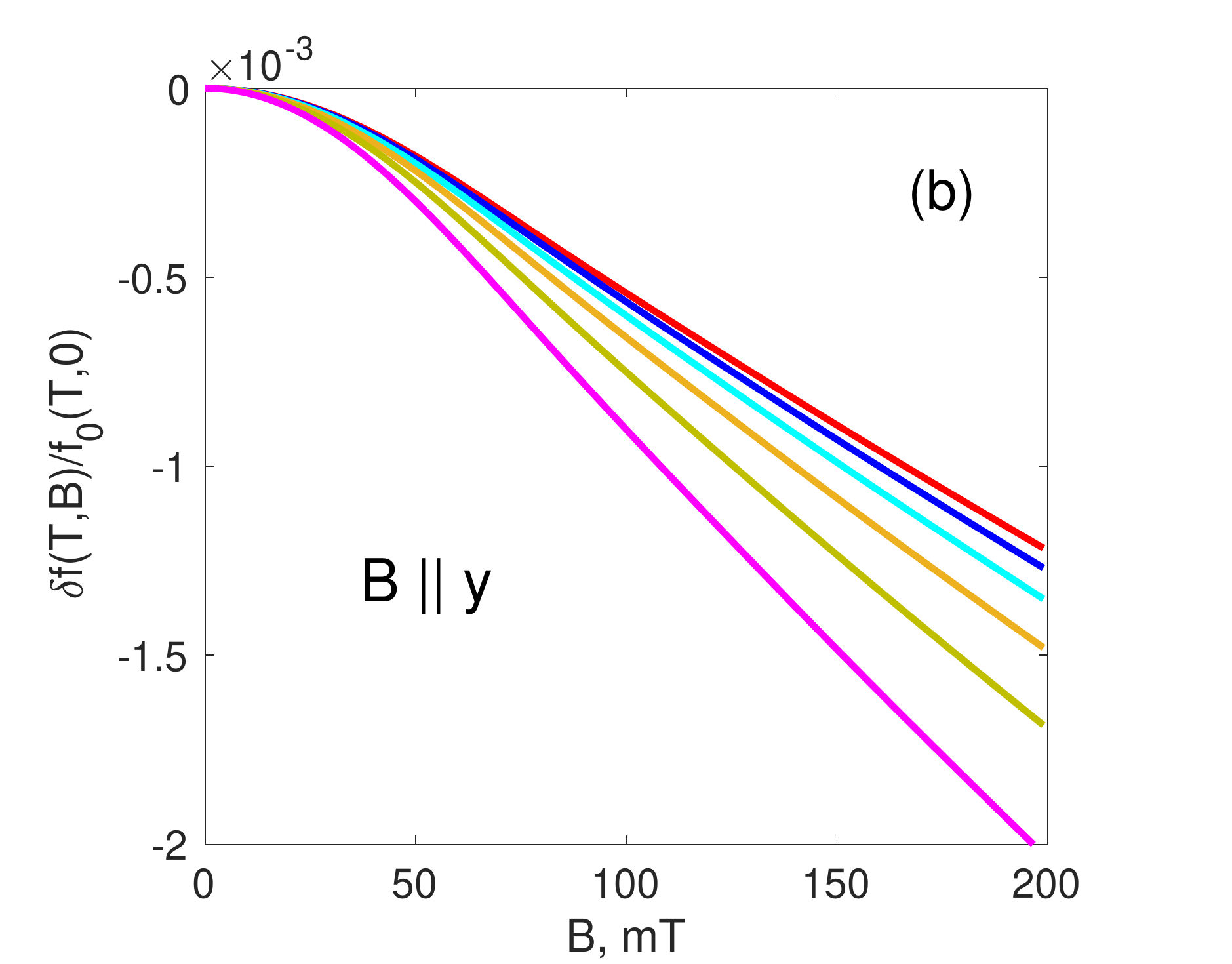}
	\caption{Normalized shift in $\delta f_0(T,B)/f_0(T,0)$ as a function of field calculated from Eqs. (\ref{dff}), (\ref{cosz}) and (\ref{cosy}) at $B_0=10$ mT, $L_k^J(0,0)/L=3.5\times 10^{-4}$ and different temperatures corresponding to those in Fig. 9 for: (a) ${\bf B}\|z$ and (b) ${\bf B}\|y$.} 
	\label{fig:11} 
\end{figure}

The fit of the observed $\delta f(T,B)/f_0(T,0)$ to the GB model depends on many uncertain parameters such as distribution of orientations and local $J_c$ values of GBs, their geometrical sizes and mechanisms of current transport through GBs.  Shown in Fig. \ref{fig:11} is an example of  $\delta f(T,B)/f_0(T,0)$ calculated for uniform distributions of the Euler angles of GBs:  
\begin{equation}
		\frac{\delta f(T,B)}{f_0(T,0)}= \left[\frac{1 + a\epsilon(T)/\langle\cos\theta(0)\rangle}{1+a\epsilon(T)/\langle\cos\theta(B)\rangle}\right]^{1/2}-1, 
		\label{dff}
\end{equation} 
where $a=L_k^J(0,0)/L$, the factor $\epsilon(T)=[1-(T/T_c)^4]^{-2}$ approximates the S-N-S temperature dependence of $J_c(T)$ in Eq. (\ref{rat}) at $7 K<T<T_c$, and 
$\langle\cos\theta (B)\rangle$ is given by Eqs. (\ref{cosz}) and (\ref{cosy}).  We took $a=3.5\times 10^{-4}$, $B_0=10$ mT and different temperatures corresponding to those in Fig. \ref{fig:hvsdff}.  As follows from Fig. \ref{fig:11}, the model captures the observed features of $\delta f(T,B)/f_0(T,0)$ for both orientations of ${\bf B}$ shown in Fig. \ref{fig:hvsdff}, although one can hardly expect a perfect fit from such a crude model. For instance, the difference between the slopes of $\delta f(T,B)/f_0(T,0)$ for two field orientations in Fig. \ref{fig:11} is about 30$\%$ higher than in Fig. \ref{fig:hvsdff}, which can occur because the GB orientations shown in Fig. \ref{fig:afm} are not completely random. Yet the GBs can dominate the field-induced frequency shift even if they only contribute less than $10^{-3}$ to the total inductance of the strip.  

Grain boundaries can also contribute to the field-dependent resistance $R_J(B)$. Indeed,  the coplanar resonator impedance per unit length, $Z_g(B)=R_J+iL_k^J$ estimated from Eq. (\ref{imp}) is given by: 
\begin{equation}
Z_g\sim \frac{i\omega \bar{L}_\square\bar{R}_\square}{wdl_2(i\omega \bar{L}_\square+\bar{R}_\square)},
\label{zz}
\end{equation}
where $\bar{L}_\square=\phi_0/2\pi J_c\langle\cos\theta\rangle$, and $\bar{R}_\square$ is the GB quasiparticle resistance per unit area averaged over the GB orientations. The field dependencies of $R_J(B)$ and $L_k^J(B)$ can be strongly affected by $\omega$ and $T$. At high frequencies, $\omega\gg \omega_c=\bar{R}_\square/\bar{L}_\square$ the impedance $Z_g\sim \bar{R}_\square/wd$ becomes independent of $B$, that is, GBs do not contribute to the NLME. For S-I-S GBs, the crossover frequency $\omega_c$ is, 
\begin{equation}
\omega_c\sim \langle\cos\theta\rangle\frac{\Delta}{\hbar}\tanh\left(\frac{\Delta}{2k_BT}\right)
\label{omc}
\end{equation}
Here $\omega_c\sim\Delta/\hbar$ if $T$ is not very close to $T_c$ and $B\lesssim B_0$, but $\omega_c(T,B)$ decreases as $B$ increases beyond $B_0$ and $T$ approaches $T_c$. For proximity-coupled GBs, $\omega_c$ can be smaller than $\Delta/\hbar$ even at $T\ll T_c$ and $B<B_0$ since  $J_cR_\square$ for S-N-S JJs can be much smaller than $\Delta/e$~ \cite{BP}. For Nb$_3$Sn in which $\Delta/\hbar\sim 1$ THz is some 2 orders of magnitude higher than the frequency $2\pi f_0$ of our resonator, it appears that the observed behavior of $\delta f(T,B)$ at $7<T<12$ K is consistent with the low-frequency limit $\omega\ll \omega_c(T,B)$ in which Eq. (\ref{zz}) gives:
\begin{equation}
R_J(B) \sim \frac{\phi_0^2\omega^2}{4\pi^2R_\square J_c^2 wdl_2\langle \cos\theta\rangle^2}
\label{rjj}
\end{equation}
For $\langle L_J(B)\rangle$ calculated above  (see Fig. \ref{Fig.Y}), $R_J(B)\propto B^2$ in Eq. (\ref{rjj}) increases quadratically with $B$ at $B\gtrsim B_0$. This field dependence is the same as for the Meissner contribution to $\sigma_\omega(B)$ in Eq. (\ref{sig}), but the orientational dependence of $R_J(B)$ is opposite to that of $\sigma_\omega(B)$. Thus, GBs can give rise to a strong nonlinearity of the electromagnetic response of polycrystalline films, which can be essential for Nb$_3$Sn thin film coatings of high-Q resonator cavities in particle accelerators ~\cite{ml,gb3,gb4,ml1}. 

In conclusion, grain boundaries and local nonstoichiometry on nanometer scales can significantly contribute to the NLME in polycrystalline Nb$_3$Sn. Particularly, GBs can cause the linear field-dependence of the magnetic penetration depth expected from a clean d-wave superconductor at low temperatures. By contrast, for elemental superconductors such as Nb ~\cite{ggc} and Al ~\cite{Al} with large coherence lengths, $\delta \lambda(T,B)\propto B^2$ is described well by the Meissner pairbreaking. However, extended crystalline defects in superconductors with short coherence lengths can radically change the field-dependence of $\delta \lambda(T,B)$, even if their contribution to the kinetic inductance at zero field is  small.  This feature can impose stringent requirements for the quality of single crystals used for the observation of manifestations of d-wave pairing in $\lambda(T,B)$, particularly in cuprates and pnictides which are prone to the weak-link behavior of grain boundaries.

\section{ACKNOWLEDGMENTS}
This work was supported by  
DOE under Grant DE-SC 100387-020.

\appendix
%%%%%%%%%
\section{Geometric inductance of the coplanar waveguide} \label{A}
The geometric inductance per unit length $L_g$ of the coplanar resonator  is calculated by conformal mapping of the cross section of the coplanar resonator into parallel plates~ \cite{collin}. For a thin film with $d\ll w$, this yields
\begin{equation} 
	L_g =\frac{\mu_0 K(k')}{4K(k)}, 
	\label{lg}
\end{equation} 
where $k=s/(2w+s)$, $k'=\sqrt{1-k^2}$, and $K(k)$ is a complete elliptic integral of the first kind. For $w = 15\,\mu\mathrm{m}$ and $s = 8.8\,\mu\mathrm{m}$, Eq. (\ref{lg}) gives $L_g=420.5$ H/m.

\section{tilted GB}\label{B}
For a rectangular JJ of width $l_1$ and length $l_2$ shown in Fig. \ref{Fig.X},  the solution of Eq. (\ref{grad}) is 
\begin{gather}
\theta=\frac{2\pi B \Lambda}{\phi_{0}}\left(n_{x}y-n_{y}x\right),\qquad{\bf B}\|z
\label{bz} \\
\theta=\frac{2\pi B\Lambda}{\phi_{0}}\left(n_{z}x-n_{x}z\right),\qquad{\bf B}\|y
\label{by}
\end{gather}
We rotate the coordinate system by the Euler angles $\alpha$ and $\beta$  about the $x$ and the $y$ axes to a new cartesian system $(X,Y,Z)$ in which $Z$ is perpendicular to the GB plane and $X$ and $Y$ are directed along the GB sides of lengths $l_1$ and $l_2$, respectively. Here $X_i=R_{ij}x_j$, where the rotation matrix $R_{ij}=R^y_{ik}R^x_{kj}$ is given by:
\begin{equation}
R_{ij}=\left(\begin{array}{ccc}
\cos\beta & 0 & \sin\beta\\
0 & 1 & 0\\
-\sin\beta & 0 & \cos\beta
\end{array}\right)\!\cdot\!\left(\begin{array}{ccc}
1 & 0 & 0\\
0 & \cos\alpha & -\sin\alpha\\
0 & \sin\alpha & \cos\alpha
\end{array}\right)\!.
\label{R}
\end{equation}
The transpose matrix $R^T_{ij}$ is then:
\begin{equation}
 R^{T}_{ij}=\left(\begin{array}{ccc}
\cos\beta\, & 0\, & -\sin\beta\\
\sin\alpha\sin\beta & \cos\alpha\, & \sin\alpha\cos\beta\\
\cos\alpha\sin\beta\, & -\sin\alpha\, & \cos\alpha\cos\beta
\end{array}\right)\!.
\label{RT}
\end{equation}
Using $x_i=R^T_{ij}X_j$ we obtain: 
\begin{gather}
x=X\cos\beta-Z\sin\beta,
\label{x} \\
y=X\sin\beta\sin\alpha+Y\cos\alpha+Z\cos\beta\sin\alpha,
\label{y}\\
z=X\sin\beta\cos\alpha-Y\sin\alpha+Z\cos\beta\cos\alpha,
\label{z} \\
\!\!\!n_x=-\sin\beta,\quad n_y=\cos\beta\sin\alpha, \quad n_z=\cos\beta\cos\alpha
\end{gather}
Equations (\ref{bz})-(\ref{by}) and (\ref{x})-(\ref{z}) with $Z=0$ give: 
\begin{gather}
\theta=-\frac{2\pi B\Lambda}{\phi_{0}}\left[X\sin\alpha+Y\sin\beta\cos\alpha\right],\quad{\bf B}\|z
\label{tz}\\
\theta=\frac{2\pi B\Lambda}{\phi_{0}}\left[X\cos\alpha-Y\sin\beta\sin\alpha\right],\quad{\bf B}\|y.
\label{ty}
\end{gather}
Next, we calculate $\bar{c}=S^{-1}\int\cos\theta(X,Y)dS$:
\begin{equation}
\bar{c} =\frac{1}{l_1l_2}\mbox{Re}\int_{-l_{1}/2}^{l_{1}/2}\!dX\int_{-l_{2}/2}^{l_{2}/2}\!dYe^{i\theta(Y,Z)},
\label{barrc}
\end{equation}
which leads to Eqs. (\ref{barc})-(\ref{qy}).

For uniform distributions of the GB orientation angles, Eqs. (\ref{cz}) and (\ref{avv}) yield at ${\bf B}\|z$:
\begin{gather}
\langle\cos\theta\rangle=
\nonumber \\
\frac{1}{\pi^2b}\int_{0}^{\pi}d\alpha\!\int_0^\pi d\beta\sin(b\cos\alpha\sin\beta)\frac{\cos\alpha\cos\beta}{\tan\beta}.
%\quad {\bf B}\|z.
\label{cosz}
\end{gather}
Likewise, we get from Eqs. (\ref{cy}) and (\ref{avv}) at ${\bf B}\|y$: 
\begin{gather}
\langle\cos\theta\rangle= 
 \nonumber \\
\frac{1}{\pi^2b}\int_{0}^{\pi}d\alpha\!\int_0^\pi d\beta\sin(b\sin\alpha\sin\beta)\frac{\cos\alpha\cos\beta}{\tan\alpha\tan\beta}.
%\quad {\bf B}\|y.
\label{cosy}
\end{gather}
At $b=B/B_0\ll 1$ we obtain from Eqs. (\ref{cosz})-(\ref{cosy}):
\begin{gather}
\langle\cos\theta\rangle=\frac{1}{4}\left(1-\frac{b^2}{32}\right),\quad {\bf B}\|z,
\label{cosz1}\\
\langle\cos\theta\rangle=\frac{1}{4}\left(1-\frac{b^2}{96}\right),\quad {\bf B}\|y.
\label{cosy1}
\end{gather}
The field ${\bf B}\| z$ applied along the strip causes stronger reduction of $\langle\cos\theta\rangle$ than the transverse field ${\bf B}\| y$.

\end{document}